\documentclass[prb,twocolumn,superscriptaddress,showpacs,amsmath,amssymb]{revtex4}
\usepackage{float}
\usepackage{amsmath}
\usepackage{bm}
\usepackage[usenames]{color}
\usepackage{verbatim}
\usepackage{graphicx}
\newcommand{\be}{\begin{equation}}
\newcommand{\ee}{\end{equation}}   
\newcommand{\bea}{\begin{eqnarray}}
\newcommand{\eea}{\end{eqnarray}}
\newcommand{\phrl}[1]{Phys.~Rev.~Lett. {\bf #1}}
\newcommand{\phrb}[1]{Phys.~Rev.~B {\bf #1}}
\newcommand{\phrx}[1]{Phys.~Rev.~X {\bf #1}}
\newcommand{\cmat}[1]{arXiv:{\bf #1}}

\newcommand{\jpcm}[1]{J.~Phys.:Condens.~Matter.{\bf #1}}
\newcommand{\bib}{\bibitem}
\newcommand{\lb}{\left[}
\newcommand{\rb}{\right]}
\newcommand{\lp}{\left(}
\newcommand{\rp}{\right)}

\renewcommand{\k}{\mathbf{k}}

\newcommand{\q}{\mathbf{q}}

\newcommand{\bC}{\bar{C}}
\newcommand{\bmu}{\bar{\mu}}
\newcommand{\bOmega}{\bar{\Omega}}
\newcommand{\tOmega}{\widetilde{\Omega}}
\newcommand{\tQ}{\widetilde{Q}}
\begin{document}

%\title{Absorptive part of AC Hall in tilted noncentrosymmetric Weyl semimetal}
%\title{Anomalous AC Hall in tilted doped Weyl semimetal with both broken time reversal and inversion symmetry}
\title{Imaginary part of Hall conductivity in tilted doped Weyl semimetal with both broken time reversal and inversion symmetry}

\author{S. P. Mukherjee}
\affiliation{Department of Physics and Astronomy, McMaster University, Hamiltion, Ontario, Canada L8S 4M1}

\author{J. P. Carbotte}
\affiliation{Department of Physics and Astronomy, McMaster University, Hamiltion, Ontario, Canada L8S 4M1}
\affiliation{Canadian Institute for Advanced Research, Toronto, Ontario, Canada M5G 1Z8}

\begin{abstract}
We consider a Weyl semimetal (WSM) with finite doping and tilt within a continuum model Hamiltonian with both broken time reversal and inversion symmetry. We calculate the absorptive part of the anomalous 
AC Hall conductivity as a function of photon energy ($\Omega$) for both type I and type II Weyl semimetal. For a given Weyl node, changing the sign of its chirality or of its tilt changes the sign of its 
contribution to the absorptive Hall conductivity with no change in magnitude. For a noncentrosymmetric system we find that there are ranges of photon energies for which only the positive or only the 
negative chirality node contributes to the imaginary (absorptive) part of the Hall conductivity. There are also other photon energies where both chirality contribute and there can be other ranges of 
$\Omega$ where there is no absorption associated with the AC Hall conductivity in type I and regions where it is instead constant for type II. We comment on implications for the absorption of circular 
polarized light.
\end{abstract}


\maketitle

\section{Introduction}
\label{sec:I}

Following the suggestion that Weyl fermions could exist in a solid state environment in the pyrochlore iridates, Rn$_{2}$Ir$_{2}$O$_{7}$,\cite{Savrasov} the proposal that they could also exist in the 
nonmagnetic noncentrosymmetric transition-metal monophosphides\cite{Dai} was soon verified experimentally for TaAs.\cite{Belopolski,Ding,Lv,Xu} Weyl nodes come in pairs of opposite chirality and are 
known to display many exotic properties\cite{Hosur} such as the existence of Fermi arcs\cite{Potter} on their surface, negative magnetoresistance,\cite{Huang,Burkov} chiral anomaly\cite{Ninomiya,Grover,Burkov1,Li} 
and anomalous Hall effect.\cite{Burkov2,Tiwari} Optical absorption\cite{Xiao} in Weyl semimetal reflect directly the dynamics of the Weyl fermions which have relativistic linear in momentum dispersion 
curves\cite{Nicol,Tabert} at low energies. It can also probe the chiral anomaly\cite{Ashby} and magneto-optics can provide further information.\cite{Carbotte} The Dirac cones which define the electronic 
dispersion curves can be tilted with respect to the energy axis. The degree of tilt defines type I (undertilted) and type II (overtilted). In case of type I the Fermi surface in the undoped material 
remains a point with the density of state zero at the Weyl node. For type II\cite{Soluyanov,Borisenko,Haubold} the cones are tipped over and there exists a finite density of state at zero energy (Weyl 
node) because of the appearance of electron and hole pockets. There have been many predictions and observations of type II Weyl among others in WTe$_{2}$\cite{Mou} and MoTe$_{2}$,\cite{Kaminski} in 
LaAlGe materials,\cite{Alidoust} in TaIrTe$_{2}$,\cite{Brink,Khim} in transition metal diphosphides\cite{Gresch} and doped materials like Mo$_{x}$W$_{1-x}$Te$_{2}$.\cite{Ishida} An extended list of 
references can be found in Ref.[\onlinecite{Haubold}]. 

The optical properties of both type I and type II Weyl semimetals have been of interest. The effect of a tilt on the AC longitudinal optical conductivity of a Weyl cone was considered by Carbotte
\cite{Carbotte1} in the clean limit and was generalized to include residual scattering.\cite{Mukherjee} The imaginary part of the anomalous AC Hall conductivity (absorptive part) was calculated for 
type I\cite{Steiner} as was the absorption of circular polarized light\cite{Mukherjee1} in tilted Weyl, both type I and II. There are optical experiments in type II Weyl\cite{Chinotti} in addition to 
type I.\cite{Xiao,Sushkov} In this paper we consider the AC anomalous absorptive Hall conductivity $\Im{\sigma_{xy}}(\Omega)$ as a function of photon energy $\Omega$ for the case of a noncentrosymmetric 
Weyl semimetal where both time reversal and inversion symmetry is broken.\cite{Shan} This was previously considered for topological insulator normal multi layered systems.\cite{Zyuzin} There are many 
other studies of tilt on the physical properties of Weyl semimetals. We mention a few here. The effect of disorder\cite{Fritz} and coulomb interactions on the tilt were considered\cite{Detassis} as was 
a possible phase transition from type I to type II Weyl.\cite{Gilbert} There is the possibility of hybrid Weyl materials where one Weyl node is type I and the other type II.\cite{Luo} The effect of tilt 
on quantum transport\cite{Trescher} and on the Nernst effect\cite{Ferreiros,Saha} has been reported.

In section II we specify our model Hamiltonian and the general formula for the AC Hall conductivity which applies to the case of both broken time reversal and inversion symmetry. Section III presents 
simple analytic results for the absorptive part of the Hall conductivity. Section IV gives numerical results for the case of type I Weyl and V for type II and discuss their implications for the absorption 
of circular polarized light. Discussion and conclusions are found in section VI.

\section{Formalism}
\label{sec:II}

We begin with the minimal continuum Hamiltion for a pair of Weyl node of opposite chirality with both time reversal and inversion symmetry broken. The first displaces the Dirac cone in momentum space by 
an amount $\pm \bf{Q}$ while the second shifts their energy by $\pm Q_{0}$. This is shown pictorially in Fig.(\ref{Fig1}). The left cone in Fig.(\ref{Fig1}a) represents a doubly degenerate Dirac cone 
with relativistic dispersion curves linear in momentum. Including broken time reversal symmetry, lifts the degeneracy and the two Weyl cones of opposite chirality no longer overlap but are shifted in 
momentum space by $\pm \bf{Q}$ (along the horizontal axis) as in the middle frame of Fig.(\ref{Fig1}a). For broken inversion symmetry the two cones are further displaced in energy by $\pm Q_{0}$ (up and 
down shift along vertical axis) in right frame of Fig.(\ref{Fig1}a). The Hamiltion is given by the following equation\cite{Shan,Zyuzin},
\be
\hat{H}_{s'}(\k)=C_{s'}(k_{z}-s' Q)+s' v \bm{\sigma}.(\k-s'Q\bm{e}_z)-s'Q_{0}
\label{Hamiltonian}
\ee
where $s'=1$ for the positive chirality Weyl node and  $s'=-1$ for negative chirality Weyl node. $C_{s'}$ describe the amount of tilting of the particular chiral node, $v$ the Fermi velocity and $\bm{e}_{i}$ the 
unit vector along the axis $x_{i}$ where $i=x, y, z$. We have taken the tilt direction to be along the z-axis in Eq.(\ref{Hamiltonian}) without loss of generality. The Pauli matrices are defined as 
usually by,
\be
\sigma_{x}=\lp\begin{array}{cc}0 & 1\\ 1 & 0 \end{array} \rp, \sigma_{y}=\lp\begin{array}{cc}0 & -\imath\\ \imath & 0 \end{array} \rp, \sigma_{z}=\lp\begin{array}{cc}1 & 0\\ 0 & -1 \end{array} \rp.
\ee
The broken inversion symmetry is introduced through the third term in the Hamiltonian. The Green's function corresponding to the above Hamiltonian is given by,
\be
G_{s'}(k,z)= \lb I_{2}z-\hat{H}_{s'}(\k)\rb^{-1},
\label{GF-definition}
\ee
where $I_{2}$ is a $2\times 2$ unit matrix. It is straight forward to show that the Green's function can be written in the following form,
\be
G_{s'}(k,\imath \omega_{n})\hspace{-0.1 cm}=\hspace{-0.2 cm}\sum_{s=\pm}\hspace{-0.15cm} \frac{1-ss'\bm{\sigma}.\bm{N}_{\k -s'Q \bm{e}_{z}}}{\imath \omega_{n}\hspace{-0.1 cm}-\hspace{-0.1 cm}C_{s'}(k_{z}-s' Q)\hspace{-0.1 cm}+\hspace{-0.1 cm}sv|\k\hspace{-0.1 cm}-\hspace{-0.1 cm}s' Q\bm{e}_z|\hspace{-0.1 cm}+\hspace{-0.1 cm}s'Q_{0}},
\label{GF-full}
\ee
where $\bm{N}_{\k-s'Q\bm{e}_{z}}=\frac{k_{x}\bm{e}_{x}+k_{y}\bm{e}_{y}+(k_{z} -s'Q)\bm{e}_{z}}{\sqrt{k^2_{x}+k^2_{y}+(k_{z}-s' Q)^2}}$ and $\omega_{n}$ is a Matsubara frequency.

Since in the subsequent sections we will discuss the behavior of the anomalous Hall conductivity $\sigma_{xy}$, we need the corresponding current-current correlation function within the realm of the 
Kubo formalism. It is defined as,
\bea
&& \hspace{-0.5 cm}\Pi_{xy}(\Omega_{m},\q)=\hspace{-0.1 cm}T\sum_{\omega_{n}}\sum_{s'=\pm}\hspace{-0.1 cm} \int \frac{d^3k}{(2\pi)^3} J_{x,s'}G_{s'}(\k+\q,\omega_{n}+\Omega_{m}) \nonumber\\
&& \times J_{y,s'}G_{s'}(\k,\omega_{n})\nonumber\\
&& =Te^2v^2\sum_{\omega_{n}}\sum_{s'=\pm} \int \frac{d^3k}{(2\pi)^3} \sigma_{x}G_{s'}(\k+\q,\omega_{n}+\Omega_{m})\times \nonumber\\
&& \sigma_{y}G_{s'}(\k,\omega_{n})
\eea
with another Matsubara frequency $\Omega_{m}$. We have used the definition of the current operators,
\be
J_{\{x,y\},s'}=s'ev\sigma_{\{x,y\}}.
\ee
With these definitions we calculate the expression for the correlation function after setting $\q$ to zero as,
\bea
&& \Pi_{xy}(\Omega_{m},0)=e^2 \sum_{s'=\pm}s'\int^{\Lambda-s'Q}_{-\Lambda-s'Q} \frac{dk_{z}}{2\pi}\int^{\infty}_{0} \frac{k_{\perp}dk_{\perp}}{2\pi}\times \nonumber \\ 
&& \biggl\{f(C_{s'}k_{z}+vk-s'Q_{0})-f(C_{s'}k_{z}-vk-s'Q_{0})\biggr\} \times \nonumber \\  
&& \frac{k_{z}}{k}\lb \frac{2v^2\Omega_{m}}{\Omega^2_{m}+4v^2k^2}\rb.
\label{Correlation-func}
\eea
Here $\Omega_{m}$ is the remaining Matsubara frequency, $\Lambda$ the cutoff, $k_{\perp}$ is the momentum perpendicular to $k_{z}$ and $f(E)=(e^{(E-\mu)/T}+1)^{-1}$ is the Fermi function at finite temperature 
$T$ with $\mu$ the chemical potential. The Matsubara frequencies need to be replaced by $\imath\Omega_{m}\to \Omega+\imath \delta$ and the conductivity is,
\bea
&& \sigma_{xy}(\Omega)=-\frac{\Pi_{xy}(\Omega,0)}{\imath\Omega}=\frac{e^2 v^2}{2\pi^2} \sum_{s'=\pm}s'\int^{\Lambda-s'Q}_{-\Lambda-s'Q} k_{z} dk_{z} \nonumber\\ 
&& \int^{\infty}_{0} \frac{k_{\perp}dk_{\perp}}{k} \biggl\{f(C_{s'}k_{z}+vk-s'Q_{0})-\nonumber\\ 
&& f(C_{s'}k_{z}-vk-s'Q_{0})\biggr\} \lb\frac{1}{4v^2k^2-\Omega^2}+\imath\pi \delta(4v^2k^2-\Omega^2)\rb.
\eea
This gives the absorptive part ($\Im{\sigma_{xy}}(\Omega)$) of the dynamic anomalous Hall optical conductivity,
\bea
&& \Im\sigma_{xy}(\Omega)=\frac{e^2v^2}{2\pi}\sum_{s'=\pm}s'\int^{\Lambda-s'Q}_{-\Lambda-s'Q} k_z dk_{z}\int^{\infty}_{0} \frac{k_{\perp}dk_{\perp}}{k} \times \nonumber \\
&& \biggl\{ f(C_{s'}k_{z}+vk-s'Q_{0})-f(C_{s'}k_{z}-vk-s'Q_{0})\biggr\}  \times \nonumber\\
&& \delta(4v^2k^2-\Omega^2).
\eea
The momentum integration over $k_{\perp}$ can be carried out using the Dirac delta function and only an integration over $k_{z}$ remains. The same Dirac delta function $\delta(4v^2k^2-\Omega^2)$ 
further limits the range of $k_{z}$ for finite photon energies and the upper and lower limits $\Lambda-s'Q$ and $-\Lambda-s'Q$ in Eq.(\ref{Correlation-func}) are replaced by $\frac{\Omega}{2v}$ and 
$-\frac{\Omega}{2v}$ respectively. Consequently the displacement of the two Weyl nodes $\pm\bf{Q}$ in momentum space drops out for large values of the momentum cut off $\Lambda$ as compared with 
$\frac{\Omega}{2v}$ as does $\Lambda$ itself taken to be large as compared with $\frac{\Omega}{v}$. We get,
\bea
&& \Im\sigma_{xy}(\Omega)=\frac{e^2v}{8\pi\Omega} \sum_{s'=\pm}s'\int^{\frac{\Omega}{2v}}_{-\frac{\Omega}{2v}} k_z dk_{z} \times \nonumber\\
&& \biggl\{ f(C_{s'}k_{z}+\frac{\Omega}{2}-s'Q_{0})-f(C_{s'}k_{z}-\frac{\Omega}{2}-s'Q_{0})\biggr\}
\label{Isigma_xy}
\eea
We will use this expression for the imaginary part of the AC Hall optical conductivity in the subsequent sections to derive the central results of  this article.

\section{Imaginary part of dynamic Hall conductivity }
\label{sec:III}

\begin{figure}[H]
\centering
\includegraphics[width=1.0in,height=1.5in, angle=0]{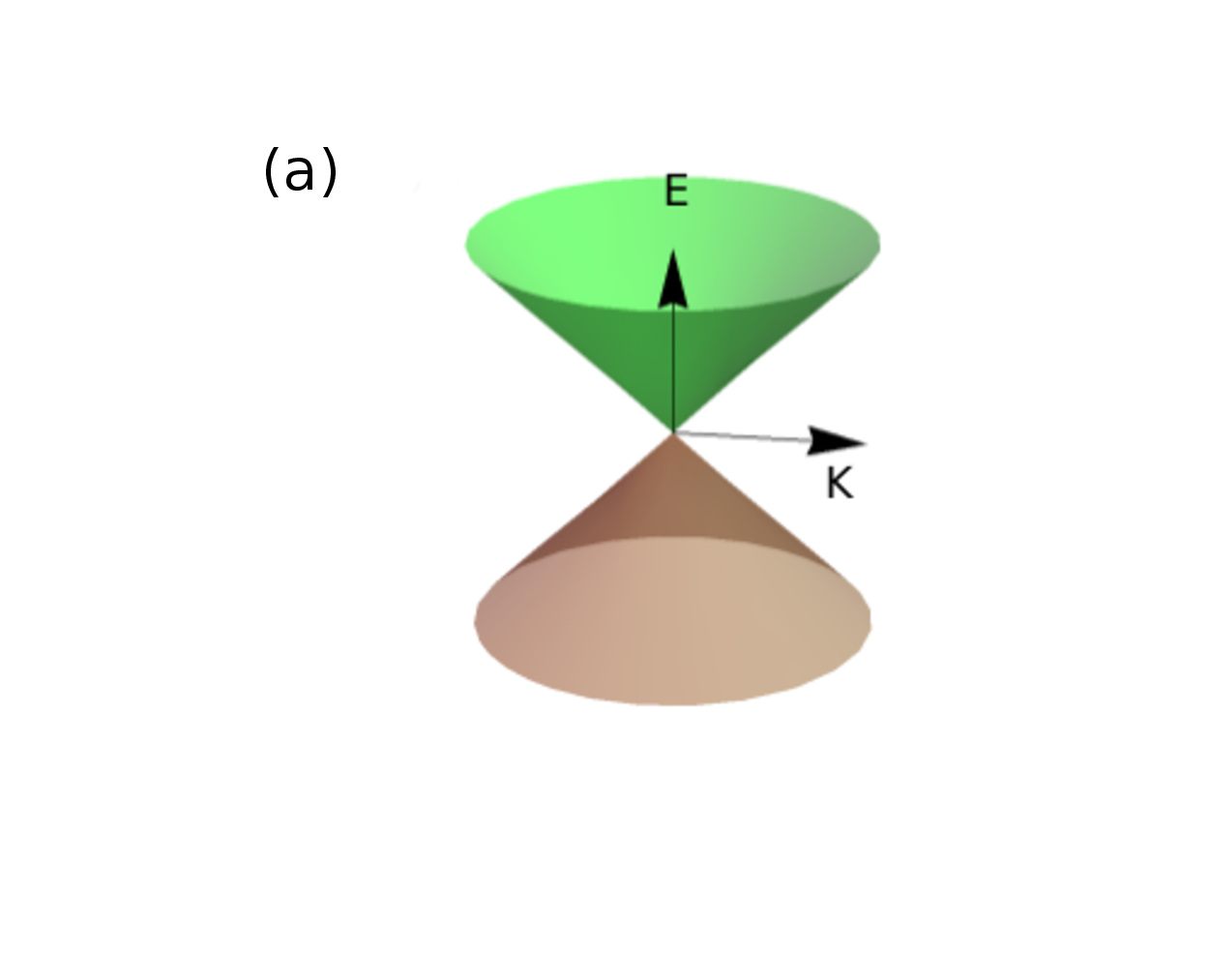}
\includegraphics[width=1.0in,height=1.5in, angle=0]{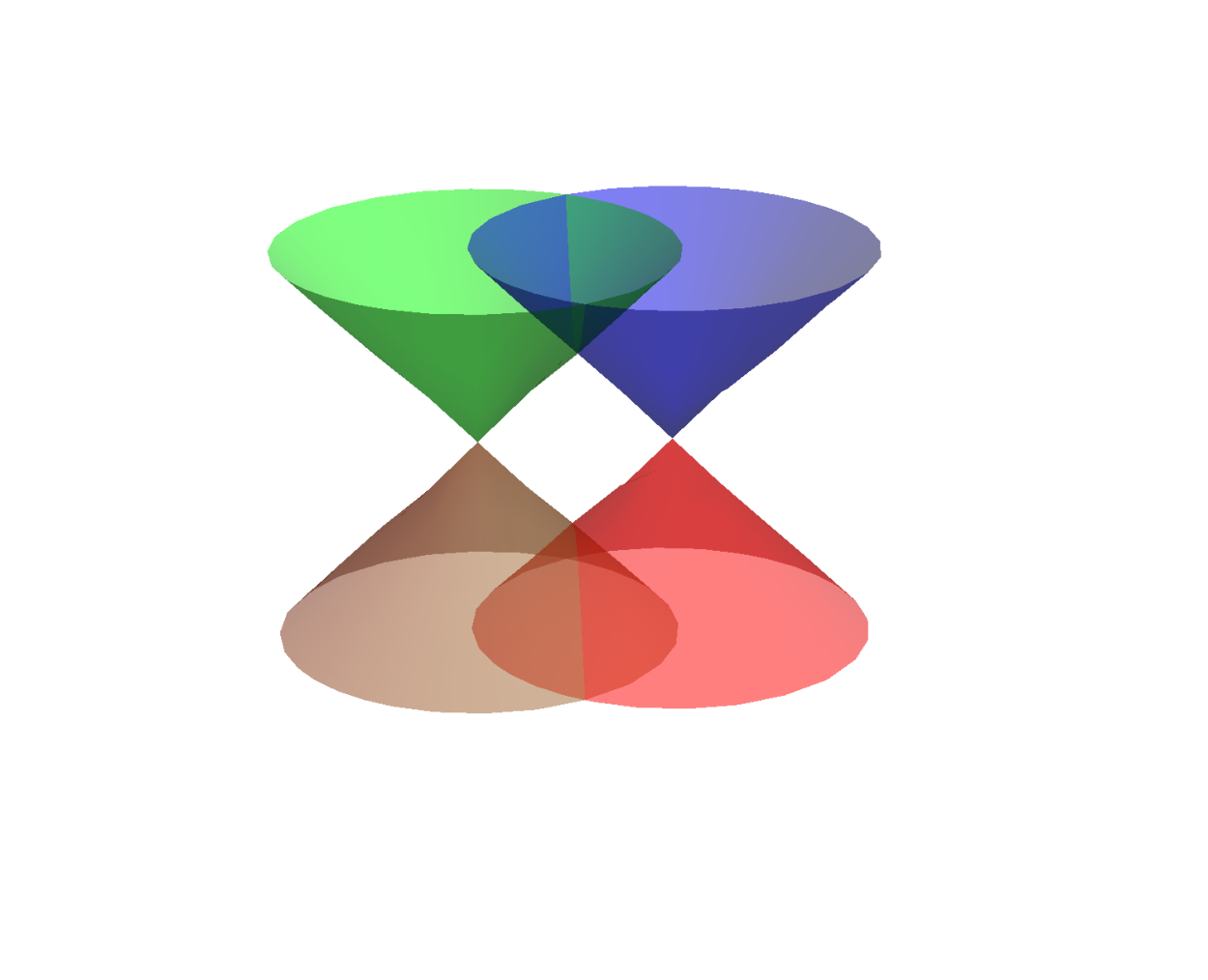}
\includegraphics[width=1.0in,height=1.5in, angle=0]{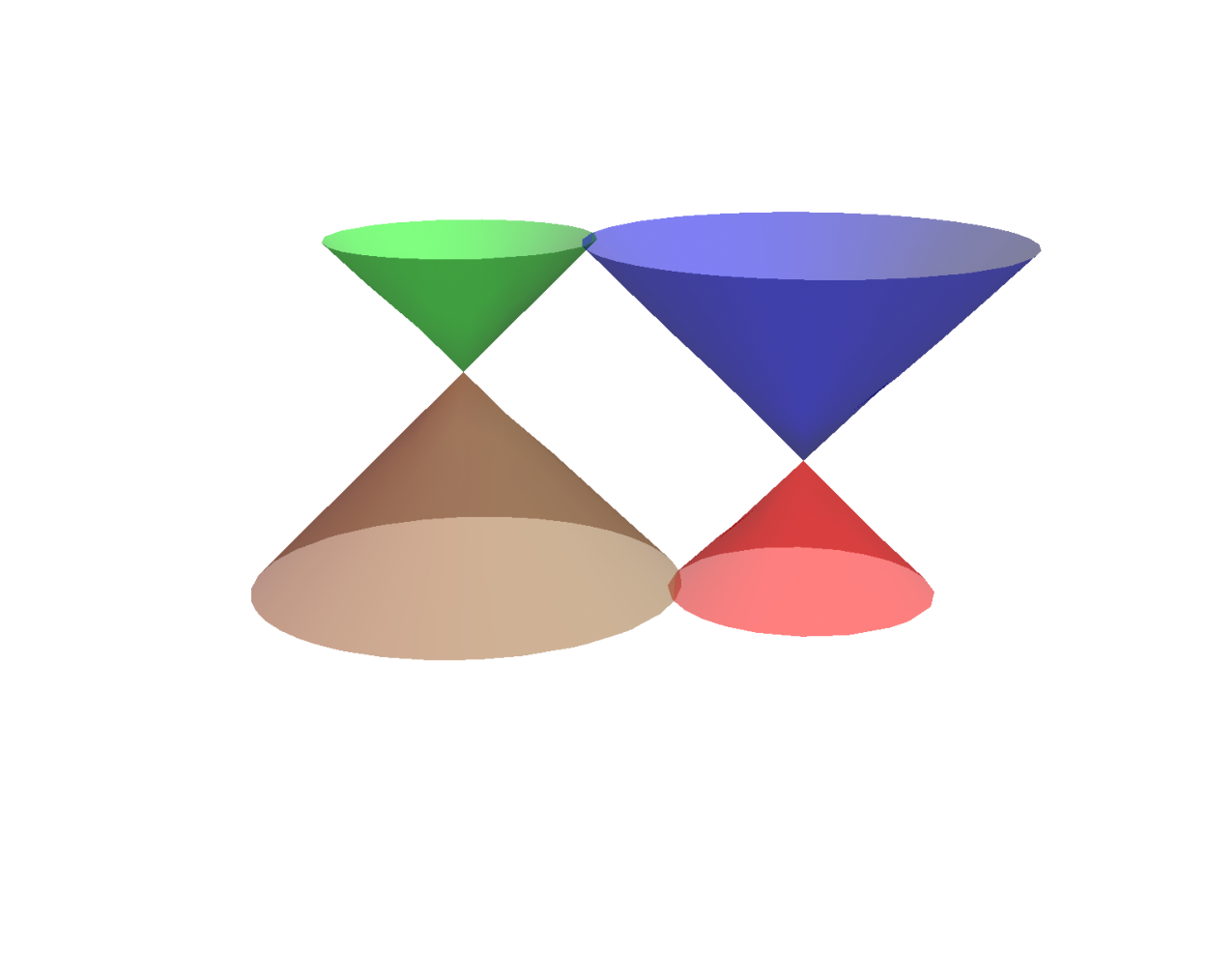}\\
\includegraphics[width=1.5in,height=1.5in, angle=0]{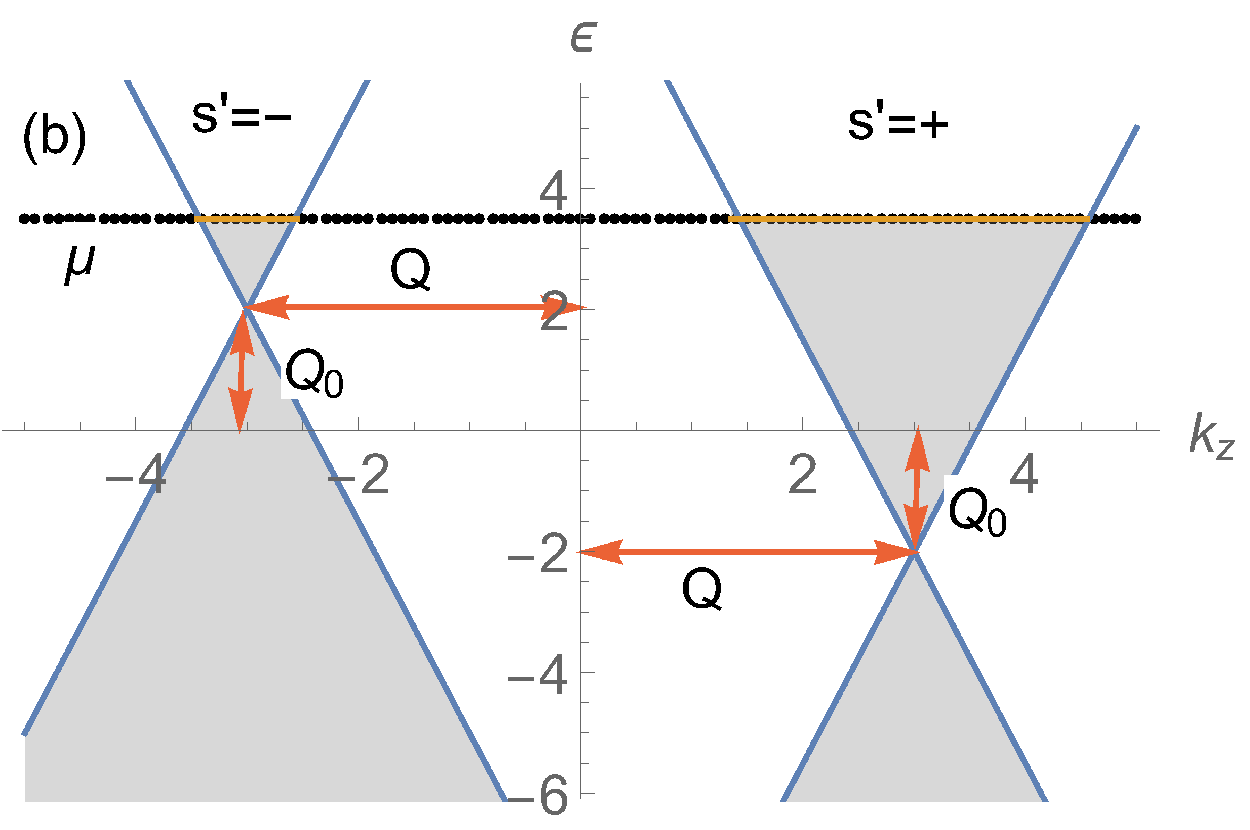}\\
\includegraphics[width=1.5in,height=1.5in, angle=0]{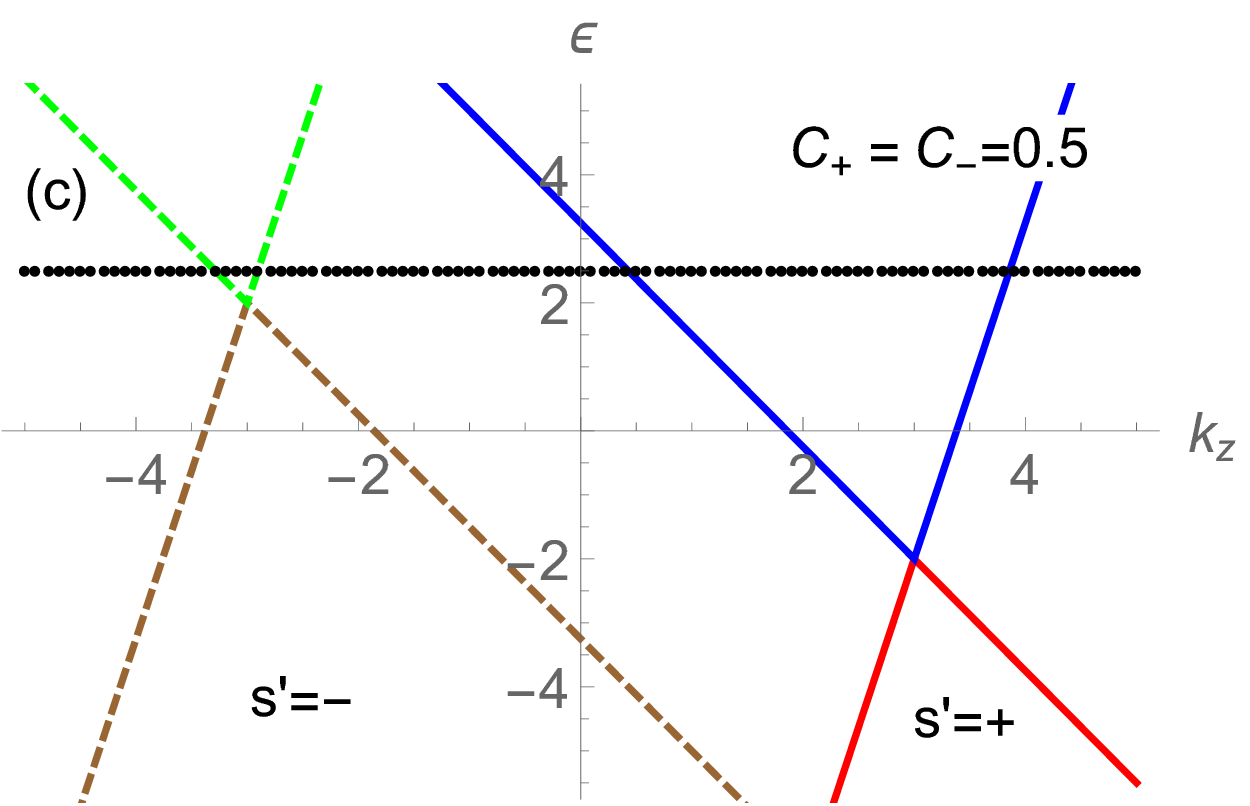}
\includegraphics[width=1.5in,height=1.5in, angle=0]{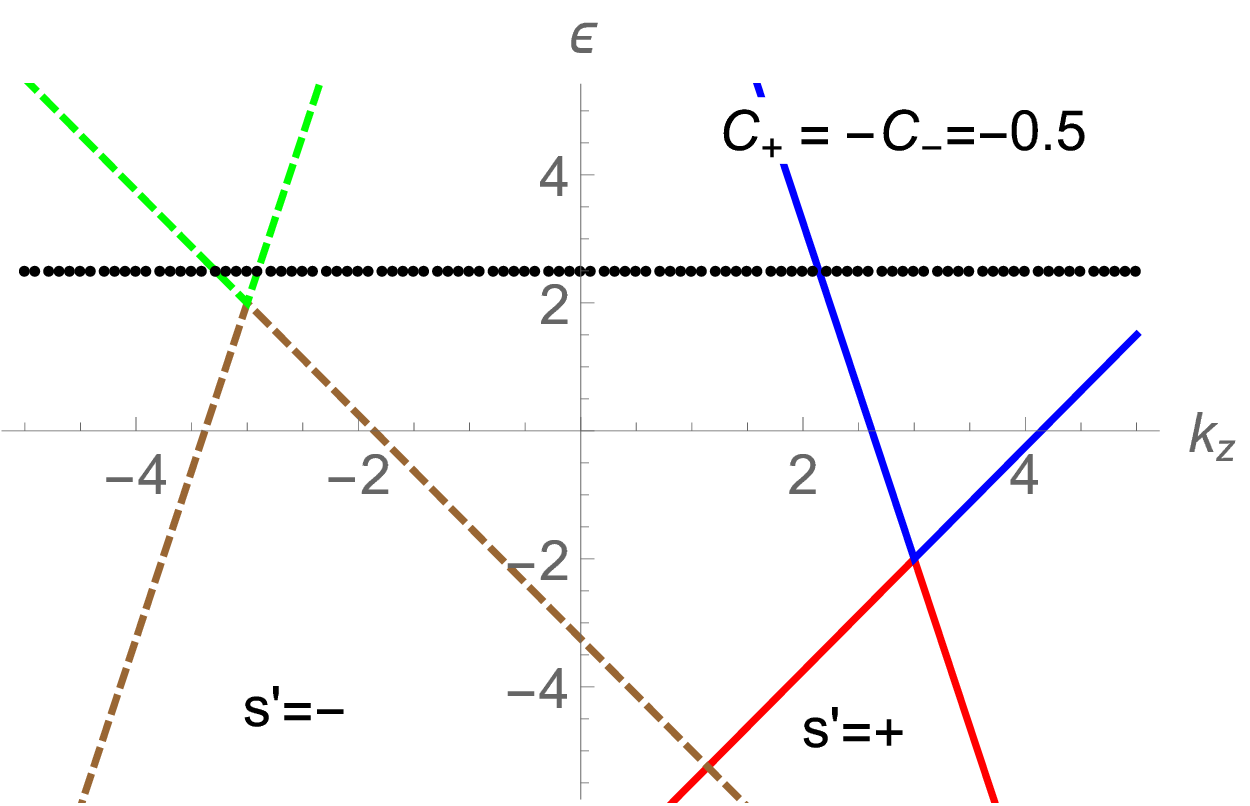}\\
\includegraphics[width=1.5in,height=1.5in, angle=0]{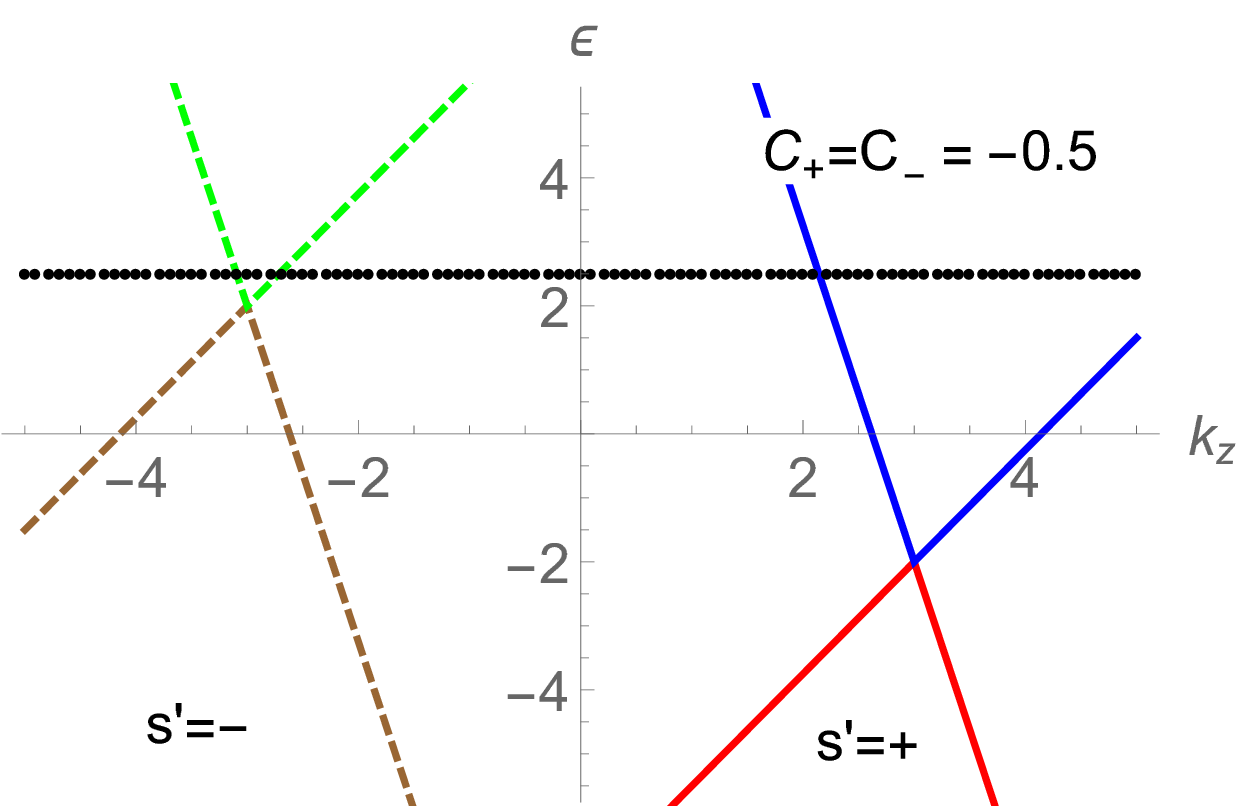}
\includegraphics[width=1.5in,height=1.5in, angle=0]{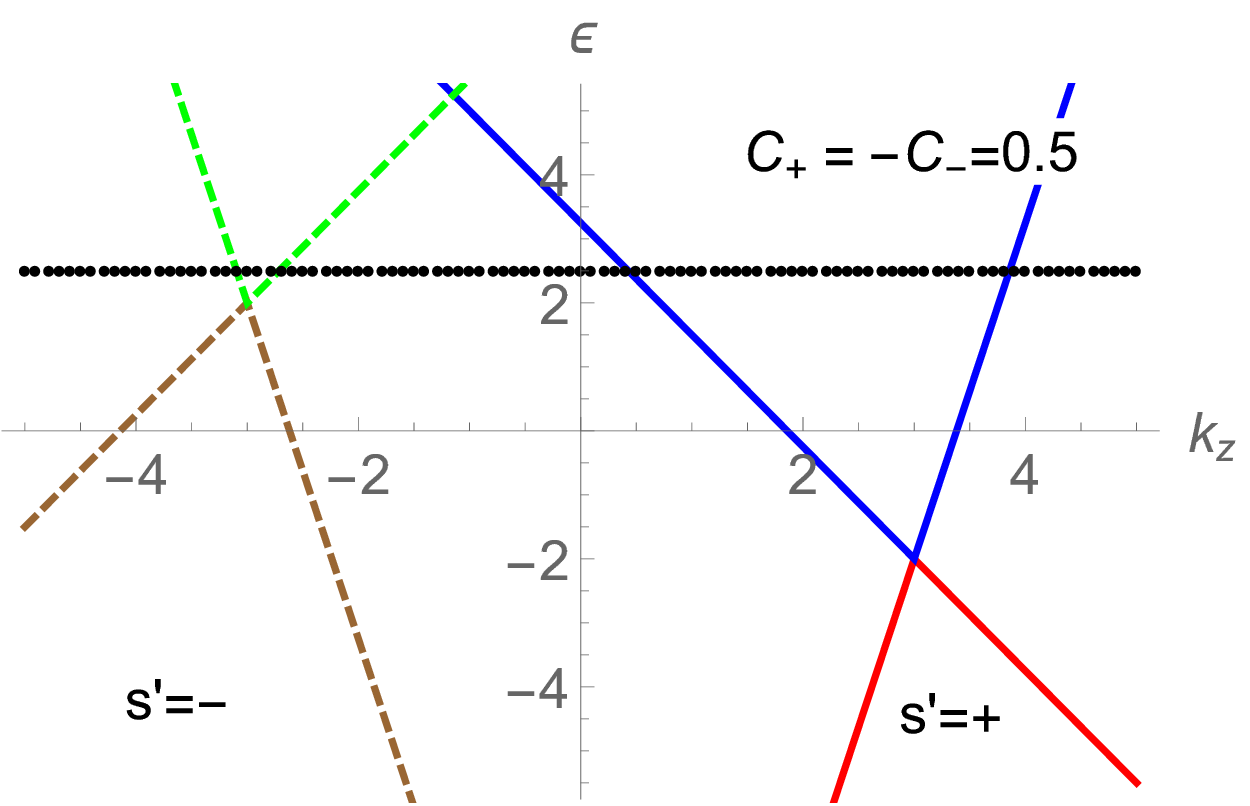}
\caption{(Color online)(a) A doubly degenerate Dirac cone (left figure) is split into two Weyl cones (middle figure) through broken time-reversal symmetry which displaces them in momentum space. A further 
breaking of inversion symmetry shifts the Weyl cones in energy relative to each other, right figure. (b) explicitly shows the momentum shift $\bf{Q}$ and energy $Q_{0}$. (c) shows possible tilt 
arrangements of the two Weyl cones. The two configurations on the right $C_{+}=-C_{-}$ oppositely tilted respect tilt inversion symmetry while the two in the left for $C_{+}=C_{-}$ (equal direction of 
the tilts) violates tilt inversion symmetry. Both tilts can be counter-clockwise to the left $C_{-}>0$ or clockwise to the right $C_{-}<0$.} 
\label{Fig1}
\end{figure}

We begin this section with the expression for the imaginary part of the Hall conductivity written in the form
\bea
&& \frac{\Im\sigma_{xy}(\Omega)}{e^2/8\pi}=\frac{1}{\Omega} \sum_{s'=\pm}s'\int^{\frac{\Omega}{2}}_{-\frac{\Omega}{2}} k_z dk_{z} \times \nonumber \\ 
&& \biggl\{ f(C_{s'}k_{z}+\frac{\Omega}{2}-\mu_{s'})-f(C_{s'}k_{z}-\frac{\Omega}{2}-\mu_{s'})\biggr\}
\label{Isigma_xy1}
\eea
where we have made the chemical potential explicit and the Fermi function is now simply $f(E)=(e^{E/T}+1)^{-1}$. We have set $v=1$ so $C_{s'}=1$ is the boundary between type I and type II. We have 
introduced separate effective chemical potentials $\mu_{s'}=\mu+s'Q_{0}$ for each Weyl nodes. Here $\mu$ is the chemical potential and charge neutrality corresponds to $\mu=0$. In the middle frame (b) of 
Fig.(\ref{Fig1}) we show a schematic of the energy of the electron dispersion curves as a function of $k_{z}$ with $\mu$ indicated by the heavy solid black line. The shaded parts of the cones are occupied 
states and we see that for the negative chirality node, the effective chemical potential is $\mu_{-}=\mu-Q_{0}$, measured from its node while the effective chemical potential for the positive chirality 
node is $\mu_{+}=\mu+Q_{0}$. It is convenient to introduce the function,
\bea
&& \hspace{-0.5cm}I(\bmu,\bC)\hspace{-0.1cm}\equiv \hspace{-0.2cm}\int^{\frac{\Omega}{2}}_{-\frac{\Omega}{2}} \hspace{-0.2cm}\frac{k_z dk_{z}}{\Omega}\hspace{-0.1cm} \biggl\{ f(\bC k_{z}\hspace{-0.1cm}+\hspace{-0.1cm}\frac{\Omega}{2}\hspace{-0.1cm}-\hspace{-0.1cm}\bmu)\hspace{-0.1cm}-\hspace{-0.1cm}f(\bC k_{z}\hspace{-0.1cm}-\hspace{-0.1cm}\frac{\Omega}{2}\hspace{-0.1cm}-\hspace{-0.1cm}\bmu)\biggr\}
\label{FuncI}
\eea
which, in our units, is the contribution of a positive chirality Weyl cone of tilt $\bC$ and chemical potential $\bmu$, to $\Im\sigma_{xy}(\Omega)$. From the form of the Hamiltonian in Eq.(\ref{Hamiltonian}) 
a positive value of tilt $C_{s'}\ge 0$ bends the Dirac cone to the left while changing the sign of $C_{s'}$ bends it to the right. This change in direction of the tilt from anticlockwise to clockwise as
$\bC$ goes from positive to negative changes the sign of the integral defined in Eq.(\ref{FuncI}). In addition, because of the explicit factor of $s'$ which appears in Eq.(\ref{Isigma_xy1}), changing the 
chirality will change the sign of this contribution to the absorptive part of the Hall conductivity. These factors will be important in what follows. In the lower frame of Fig.(\ref{Fig1}) we show the 
electronic dispersion curves for the four arrangements of tilt in the specific case of $C_{-}=0.5$. In the top left frame both $C_{+}$ and $C_{-}$ are positive and both cones are tilted counterclockwise 
while in the bottom left $C_{+}$ and $C_{-}$ are negative and both tilts are clockwise. In the right column top frame the negative chirality node is tilted counterclockwise and the positive chirality 
clockwise. For the lower frame (right column) the opposite holds. These different orientations of tilt affect the anomalous Hall conductivity. Recalling that changing the sign of the chemical potential 
gives $-I(|\bmu|,\bC)$ in Eq.(\ref{FuncI}) and working out the integral gives for $\bC<1$,
\bea
&& I(\bmu,\bC)=\text{sign}(\bmu)\lb\frac{1}{8}(\frac{1}{\bC^2}-1)\Omega-\frac{|\bmu|}{2\bC^2}+\frac{|\bmu|^2}{2\bC^2 \Omega}\rb\nonumber\\
&& \hspace{3.0cm}~~\text{for}~~\Omega_{L}<\Omega<\Omega_{U}
\label{FuncIcmu}
\eea
and zero otherwise. Here $\Omega_{L,U}=\frac{2|\bmu|}{|1\pm\bC|}$ and we have built in the symmetry between $\bmu$ and $-\bmu$. On the other hand, for $\bC>1$, $I(\bmu,\bC)$ is same as in Eq.(\ref{FuncIcmu}) 
for $\Omega_{L}<\Omega<\Omega_{U}$. For $\Omega>\Omega_{U}$ the integral in Eq.(\ref{FuncIcmu}) is no longer zero and instead is equal to $J(\bmu,\bC)$ with,

\be
J(\bmu,\bC)=-\frac{\text{sign}(\bmu)|\bmu|}{\bC^2} ~~\text{for}~~\Omega>\Omega_{U}.
\label{FuncIcmu1}
\ee
These functions play a fundamental role in this work. In our case we have two Weyl nodes. The positive (negative) chirality node has effective chemical potential $\mu_{\pm}=\mu\pm Q_{0}$ as shown in
Fig.(\ref{Fig1}b). We note that of these two effective chemical potentials $\mu_{-}$ can be positive or negative depending on the relative magnitude of $\mu$ and $Q_{0}$ ($\mu>Q_{0}$ or $\mu<Q_{0}$) 
while $\mu_{+}$ is always positive.
\begin{figure}[H]
\centering
\includegraphics[width=1.5in,height=1.5in, angle=0]{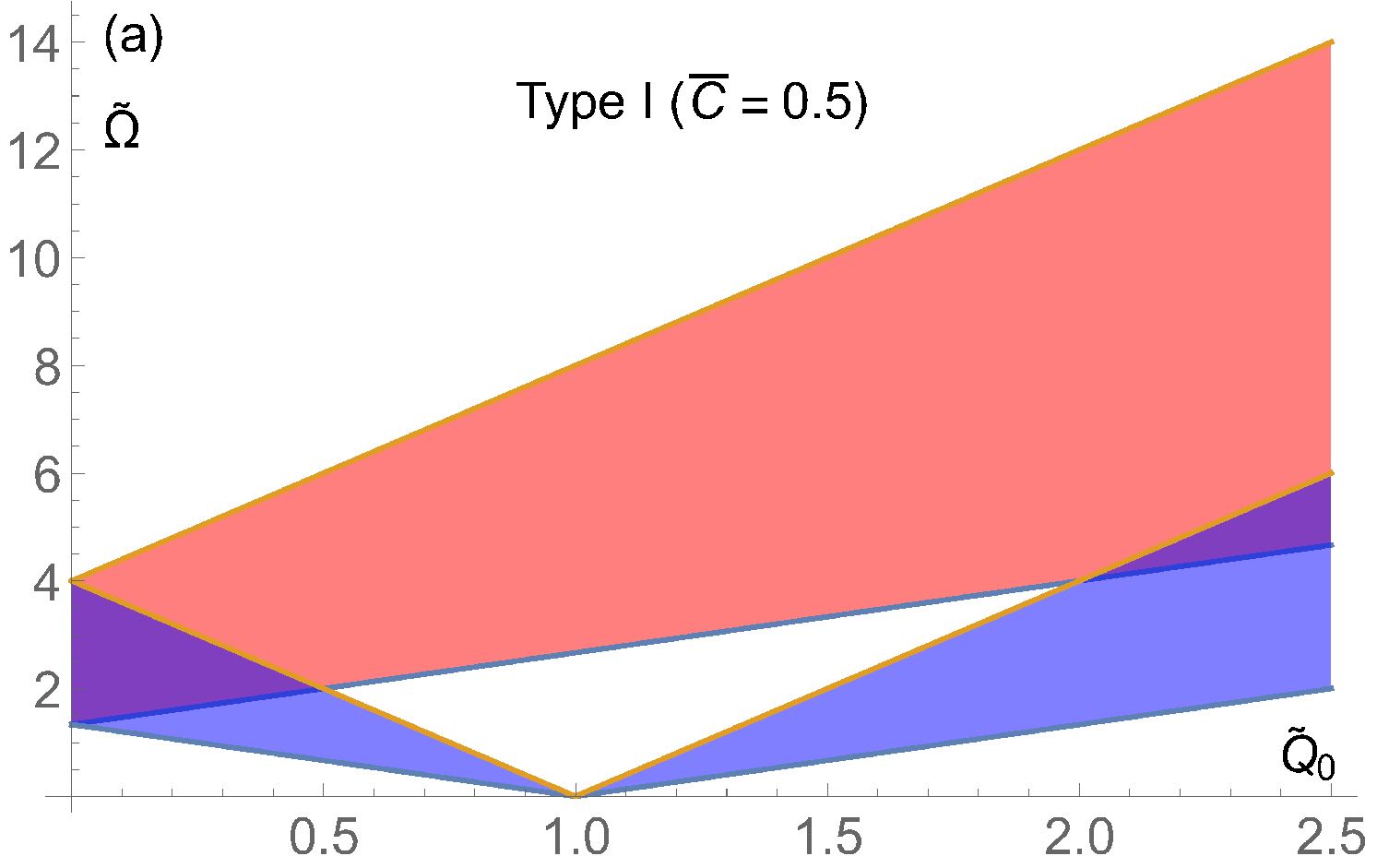}
\includegraphics[width=1.5in,height=1.5in, angle=0]{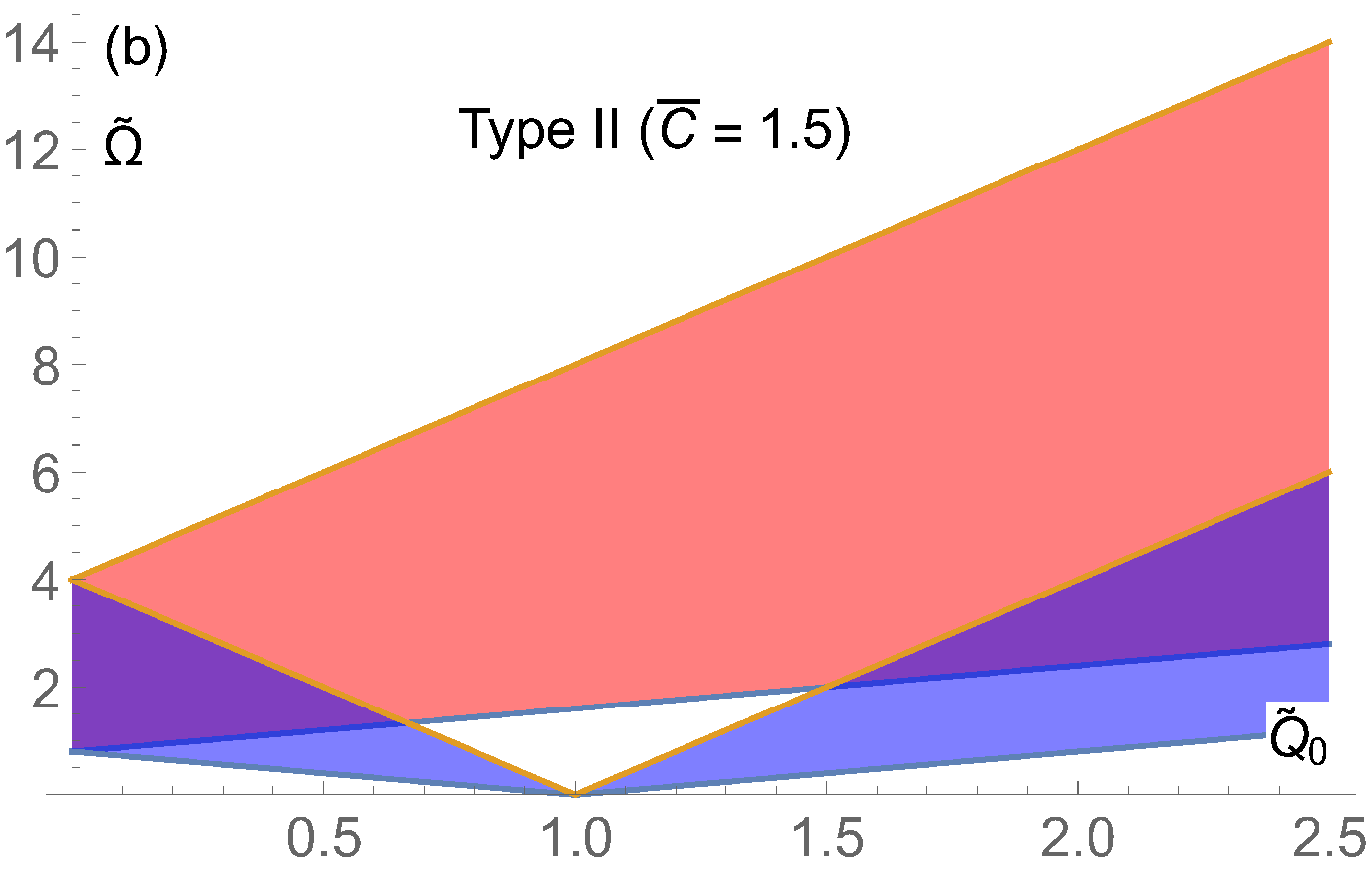}
\caption{(Color online) Frame (a) applies to type I, point Fermi surface at zero doping. The white regions are photon energies at which the absorptive Hall conductivity vanishes, this includes the entire 
region above the red shaded region. The pure blue regions involve only the negative chirality node while the pure red involves solely the positive chirality. The overlap regions have contributions from 
both nodes. Note that for $\tQ_{0}=0$, centrosymmetric case the two nodes never contribute separately. It is the presence of a finite $\tQ_{0}$ which separates these contributions. (b) applies to type II 
Weyl, where Fermi surface at zero doping involves electron and hole pockets. The lined regions over the white background are regions of photon energy where the Hall absorption is constant. In the lower 
part of the figure only the negative node contributes. At the top of the figure they both contribute a constant piece.} 
\label{Fig2}
\end{figure}
Introducing the ratio $\tQ_{0}=Q_{0}/\mu$, the limits involved in Eq.(\ref{FuncIcmu}) and Eq.(\ref{FuncIcmu1}) are,
\be
\tOmega^{+}_{L,U}=2\biggl|\frac{1+\tQ_{0}}{1\pm\bC}\biggr|,\tOmega^{-}_{L,U}=2\biggl|\frac{1-\tQ_{0}}{1\pm\bC}\biggr|
\ee
where $\tOmega^{+}_{L,U}$ is a lower(upper) limit for the positive chirality node and $\tOmega^{-}_{L,U}$ is the equivalent pair for the negative chirality Weyl node. In Fig.(\ref{Fig2}) we show these 
boundaries in a plot of $\tOmega$ vs $\tQ_{0}$. Frame (a) applies for $\bC<1$ (type I) with $\bC=0.5$ chosen for definiteness and frame (b) is for $\bC>1$ (type II) with $\bC=1.5$. At $\tQ_{0}=0$ 
(centrosymmetric) for type I, $\tOmega^{+}_{L}$ and $\tOmega^{-}_{L}$ merge at $\tOmega=4$ and 1.34 respectively. At $\tQ_{0}=1$ the boundary at which the effective chemical potential of the negative 
chirality node changes sign $\tOmega^{-}_{L}=\tOmega^{-}_{U}=0$. There is a crossing of $\tOmega^{+}_{L}=\tOmega^{-}_{U}$ at $\tQ_{0}=\bC$, here 0.5 with a second crossing at $\tQ_{0}=1/\bC$, here 2. For 
type II at $\tQ_{0}=0$ (centrosymmetric) $\tOmega^{+}_{L}=\tOmega^{-}_{L}=2/(1+\bC)=0.8$ while $\tOmega^{+}_{U}=\tOmega^{-}_{U}$ remains at 4, $\tOmega^{-}_{U}=\tOmega^{-}_{L}=0$ at $\tQ_{0}=1$ but 
$\tOmega^{+}_{L}$ crosses with $\tOmega^{-}_{U}$ at $\tQ_{0}=1/\bC$ equal to 0.67 here and again at $\bC$ equal to 1.5 in this example. In the phase diagram of Fig.(\ref{Fig2}) for $\tOmega$ as a function
of $\tQ_{0}$ the shaded blue region are bounded by $\tOmega^{-}_{L}$ (at bottom) and $\tOmega^{-}_{U}$ (at top) while the red regions are bounded by $\tOmega^{+}_{L}$ (at bottom) and $\tOmega^{+}_{U}$ 
(at top). The blue region involves the negative chirality node only while the red involves only the positive chirality node but there are overlap regions where both contribute. In particular for the 
centrosymmetric case ($\tQ_{0}=0$) the overlap is complete and both nodes contribute equally. As $\tQ_{0}$ increases the overlap region decreases and there is a low region where only the negative chirality 
node contributes and a second region at higher $\tOmega$ where only the positive node contributes. For type I the overlap region ends at $\tQ_{0}=\bC$ and in the range $\tQ_{0}=\bC$ to $\tQ_{0}=1/\bC$ the 
nodes of opposite chirality contribute to separate ranges of photon energies with no overlap. At $\tQ_{0}=1$, $\tOmega^{-}_{L}$ goes through zero because the effective chemical potential of the negative 
chirality node has vanished. As we have noted this change in sign of the chemical potential changes the sign of this contribution to the anomalous Hall (imaginary part) but not its magnitude. Beyond 
$\tQ_{0}=2$ a second overlap region is seen but there remains a large range of photon energies at low $\tOmega$ which involves only the $s'=-1$ node and another at high $\tOmega$ which is due only to the 
$s'=+1$ node. Finally we note that between $\tQ_{0}=\bC$ and $1/\bC$ there is a region where there is no absorption at all between $\tOmega=\tOmega^{-}_{U}$ and $\tOmega=\tOmega^{+}_{L}$. For type II right 
frame in Fig.(\ref{Fig2}) the overall qualitative picture remains similar with two differences that need to be emphasized. The crossing of $\tOmega^{+}_{L}$ with $\tOmega^{-}_{U}$ are now at $\tQ_{0}$ 
equal to $1/\bC$ and $\bC$ respectively and the unshaded region bounded between the red and blue regions (Fig.(\ref{Fig2}b)) in this range of $\tQ_{0}$ no longer corresponds to zero absorptive Hall 
conductivity. Instead there remains a contribution from the negative chirality node because this contribution no longer becomes zero above $\tOmega^{-}_{U}$ but remains finite to large values of photon 
energies. From all this information and the integrals defined  in Eq.(\ref{FuncIcmu}) and (\ref{FuncIcmu1}) we can calculate explicitly the imaginary part of the anomalous Hall conductivity.

\section{Results for type I}
\label{sec:IV}

\begin{figure}[H]
\centering
\includegraphics[width=2.0in,height=2.0in, angle=0]{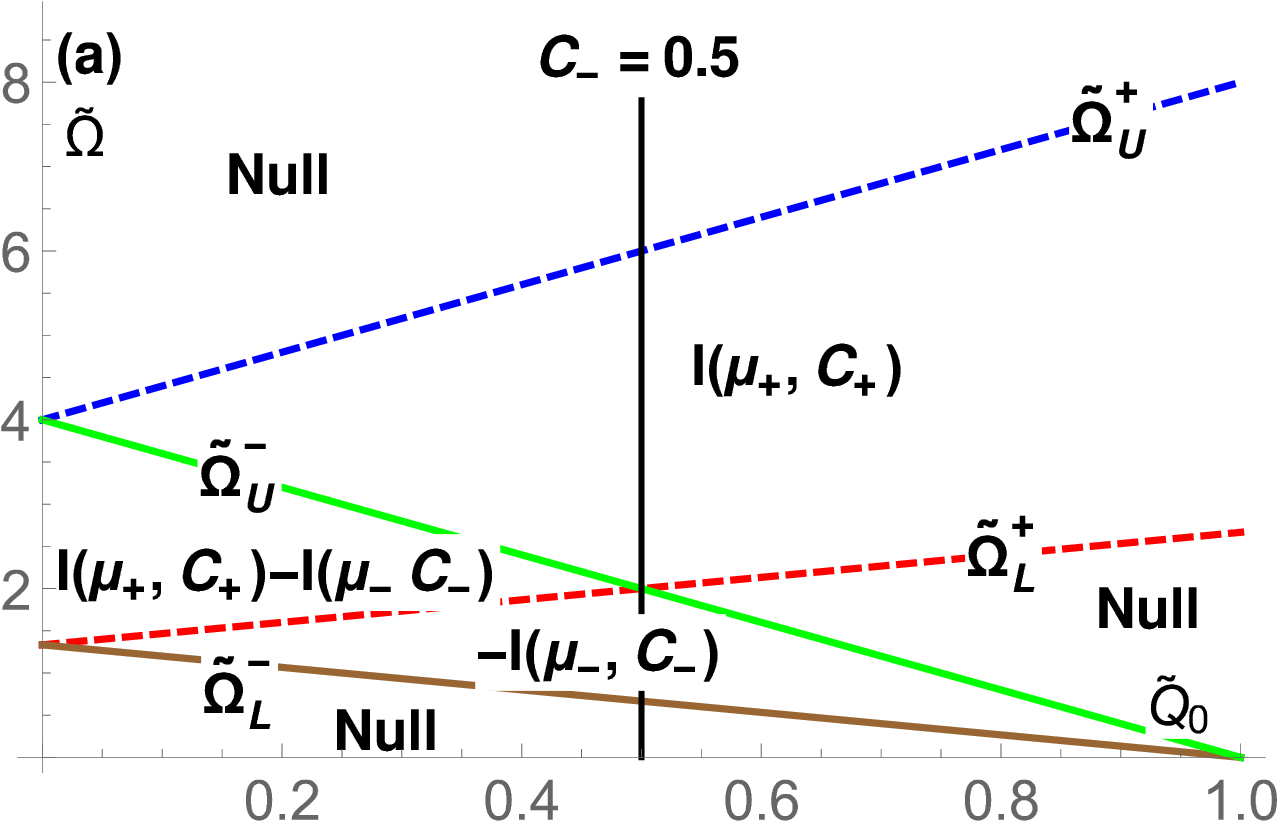}
\includegraphics[width=2.0in,height=2.5in, angle=270]{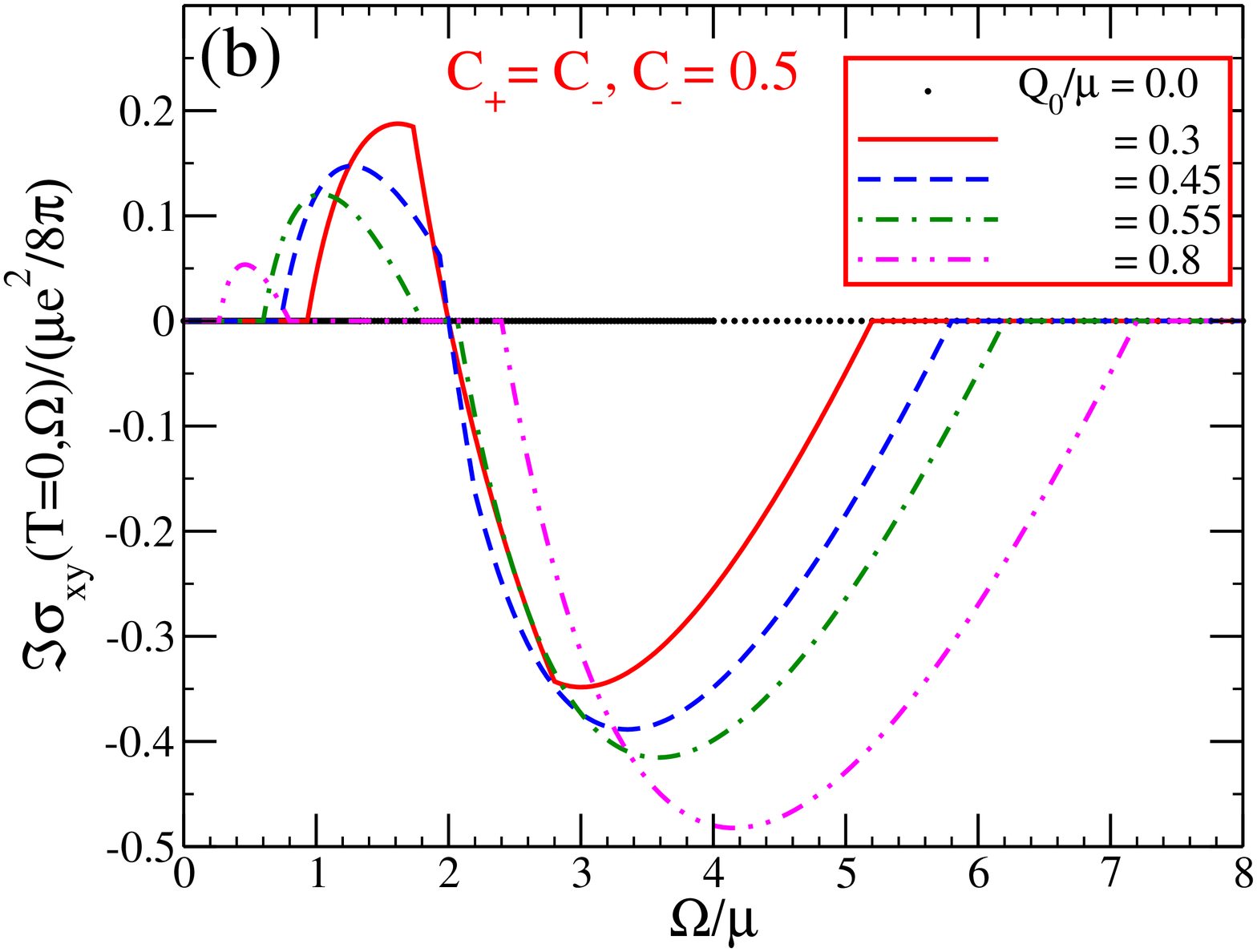}
\includegraphics[width=2.0in,height=2.5in, angle=270]{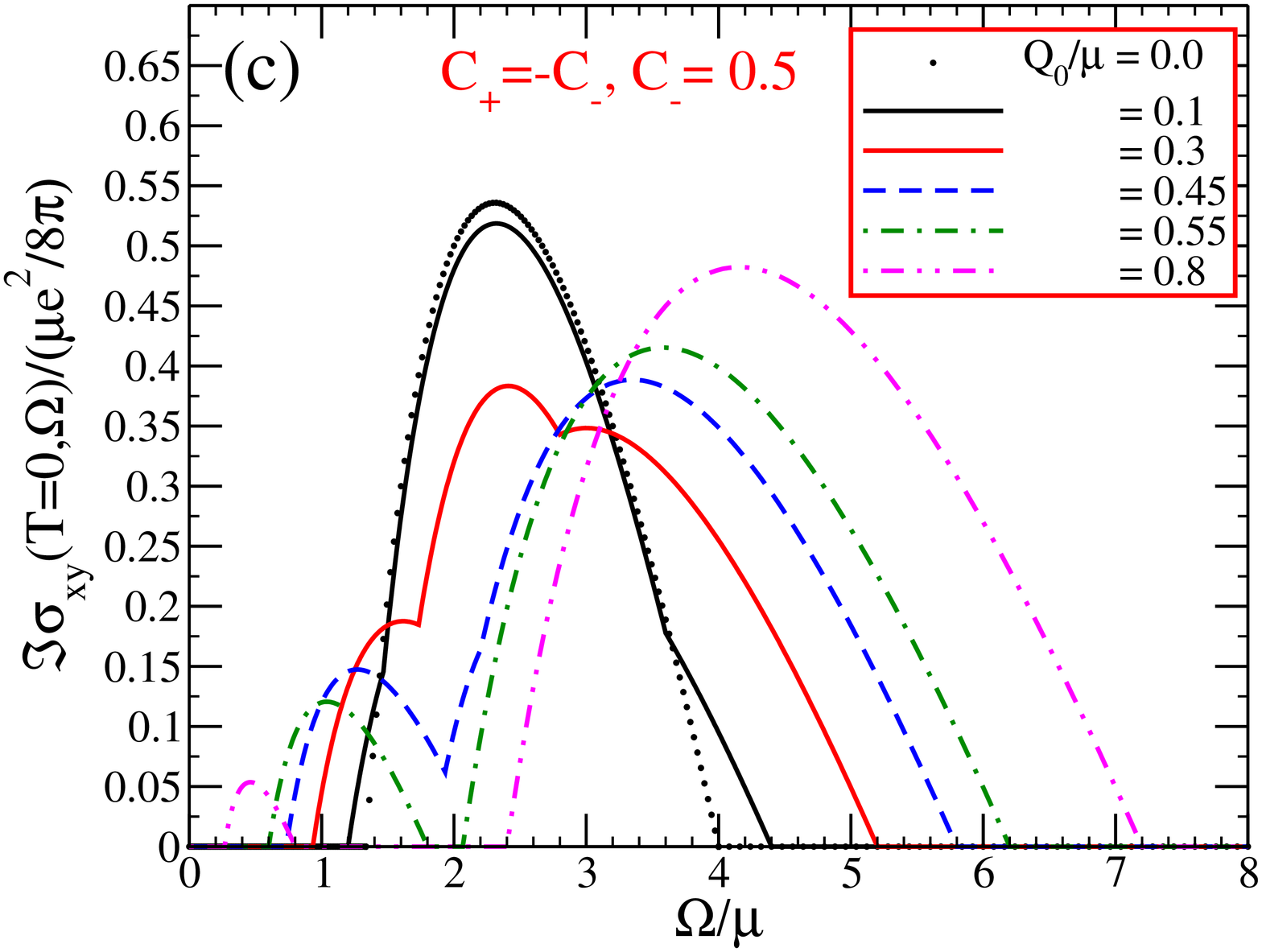}
\caption{(Color online) (a) Extended version of the phase diagram of Fig.(\ref{Fig2}a) for $\tQ_{0}<1$ with the contribution to the absorptive Hall conductivity indicated in the various distinct regions 
of the ($\tOmega,\tQ_{0}$) plane. `Null' indicates the regions of no absorptive Hall. The value of both the tilts are assumed to be $0.5$. In the other two sub-figures we plot the imaginary (absorptive) 
part of the AC Hall conductivity in units of $e^2 \mu/8\pi$ as a function of photon energy $\tOmega$, normalized to the chemical potential. This normalization means that the curves are universal, dependent 
only on the parameter $\tQ_{0}=Q_{0}/\mu$. Here $C_{-}=0.5$ for definiteness and $\tQ_{0}<1$. In frame (b) $C_{+}=C_{-}$ while in frame (c) $C_{+}=-C_{-}$. This change in tilt from counterclockwise to 
clockwise of the positive chirality cone changes the sign of its contribution to $\Im{\sigma_{xy}}(\Omega)$.} 
\label{Fig3}
\end{figure}

We begin with the case $0<C_{-}<1$ type I Weyl and consider first the case when $\tQ_{0}<1$ for which both $\mu_{+}$ and $\mu_{-}$ are always positive. In Fig.(\ref{Fig3}) frame (a)(top) we redraw that 
part of the phase diagram and indicate the value the imaginary part has, in units of $e^2\mu/8\pi$, in the various regions. Only the integral $I(\mu_{+},C_{+})$ and $I(\mu_{-},C_{-})$ are needed. The 
absorptive AC Hall conductivity is zero below the line defined by $\tOmega^{-}_{L}$ and above the line defined by $\tOmega^{+}_{U}$. It is also zero in the region $0.5<\tQ_{0}<1$ between $\tOmega^{-}_{U}$ 
and $\tOmega^{+}_{L}$ so that there is no overlap of the contributions from negative and positive chirality nodes and these are separated with a region of no absorption. The negative chirality gives the 
low energy piece while the positive chirality node gives the higher energy part. The range of the zero absorption region vanishes at $\tQ_{0}=C_{-}=0.5$ and increases as $\tQ_{0}$ is increased so that 
the separation between positive and negative chirality regions increases. For $\tQ_{0}<C_{-}=0.5$ the phase diagram is more complex and there is a overlap between the positive and negative chirality 
contributions. The region between  $\tOmega^{-}_{L}$ and $\tOmega^{+}_{L}$ is entirely due to the negative chirality node and between $\tOmega^{-}_{U}$ and $\tOmega^{+}_{U}$ to the positive chirality. 
Between  $\tOmega^{+}_{L}$ and $\tOmega^{-}_{U}$ there is overlap of these two contributions. The overlap is complete at $\tQ_{0}=0$. Since we are considering the case $C_{+}=C_{-}$ we get zero in this 
limit as seen in Fig.(\ref{Fig3}b) where we show our results for $\Im{\sigma_{xy}}(\Omega)$ vs $\tOmega\equiv \Omega/\mu$. The black dots apply and they fall on horizontal axis. This cancellation can 
be traced to the fact that the chirality provides a factor of $s'$ in Eq.(\ref{Isigma_xy1}). If however we had chosen the case $C_{+}=-C_{-}$ the sign of the contribution of the positive chirality would 
be changed and so the two contributions would add instead of canceling as seen in Fig.(\ref{Fig3}) frame (c)(see black dots and solid black curve). This also holds for other values of $\tQ_{0}$ where 
positive and negative chirality contributions interfere destructively when $C_{+}=C_{-}$ and add when $C_{+}=-C_{-}$. We note that dash double dot magenta ($\tQ_{0}=0.8$) curve and dash dot green 
($\tQ_{0}=0.55$) curves in Fig.(\ref{Fig3}b) show an intermediate region of zero ($\Im\sigma_{xy}(\Omega)=0$) because they fall in the no overlap region $\tQ_{0}>C_{-}=0.5$ while the others all overlap. 
The region of no overlap is particularly interesting when the absorption of circular polarized light is considered. The conductivity for right ($+$) and left ($-$) hand circular polarization ($\sigma_{\pm}$) 
is given by \cite{Mukherjee1} the real part of the longitudinal conductivity $\Re\sigma_{xx}(\Omega)$ to which we add $\mp\Im\sigma_{xy}(\Omega)$. In the smaller $\tOmega$ region 
$\sigma_{+}<\sigma_{-}$, in the overlap region $\sigma_{+}=\sigma_{-}$ and in the third region at higher photon energy $\sigma_{+}>\sigma_{-}$ for the parallel tilted case and $\sigma_{+}<\sigma_{-}$ 
when the tilts are oppositely directed. We emphasize that increasing $\tQ_{0}$ to be greater than $C_{-}$ has completely split the negative and positive Weyl node contributions and opened a range of photon 
energy between them where the polarization of the light does not change the absorption. 

\begin{figure}[H]
\centering
\includegraphics[width=2.0in,height=2.0in, angle=0]{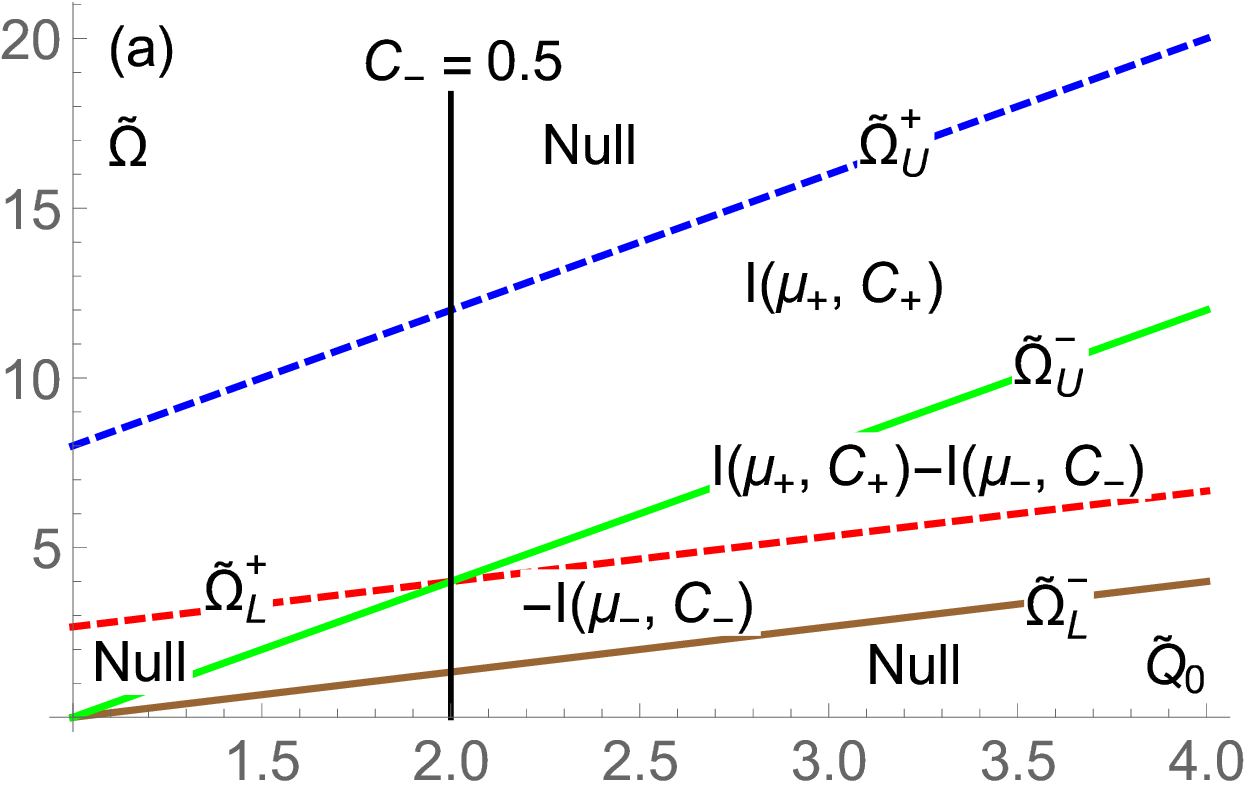}
\includegraphics[width=2.0in,height=2.5in, angle=270]{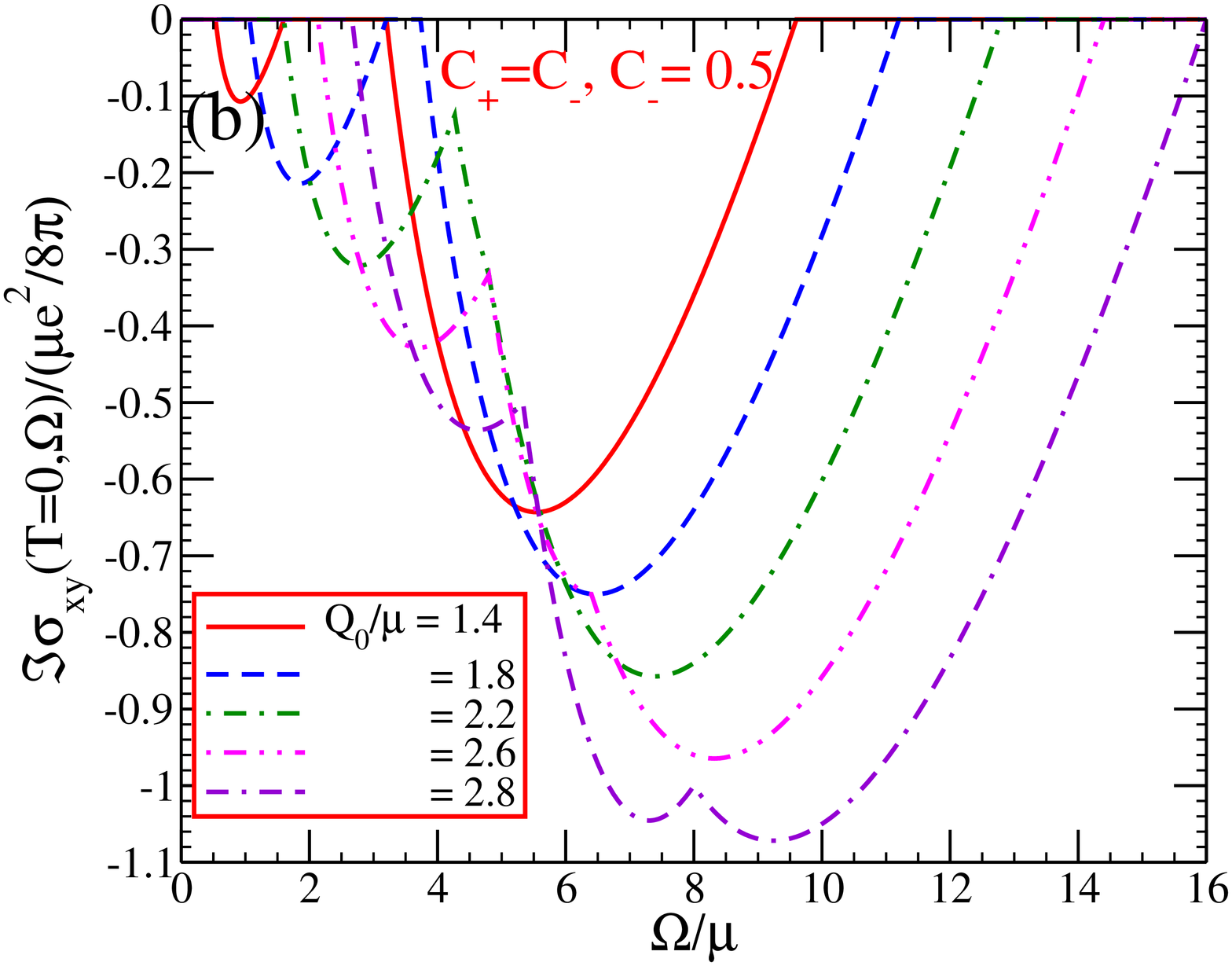}
\includegraphics[width=2.0in,height=2.5in, angle=270]{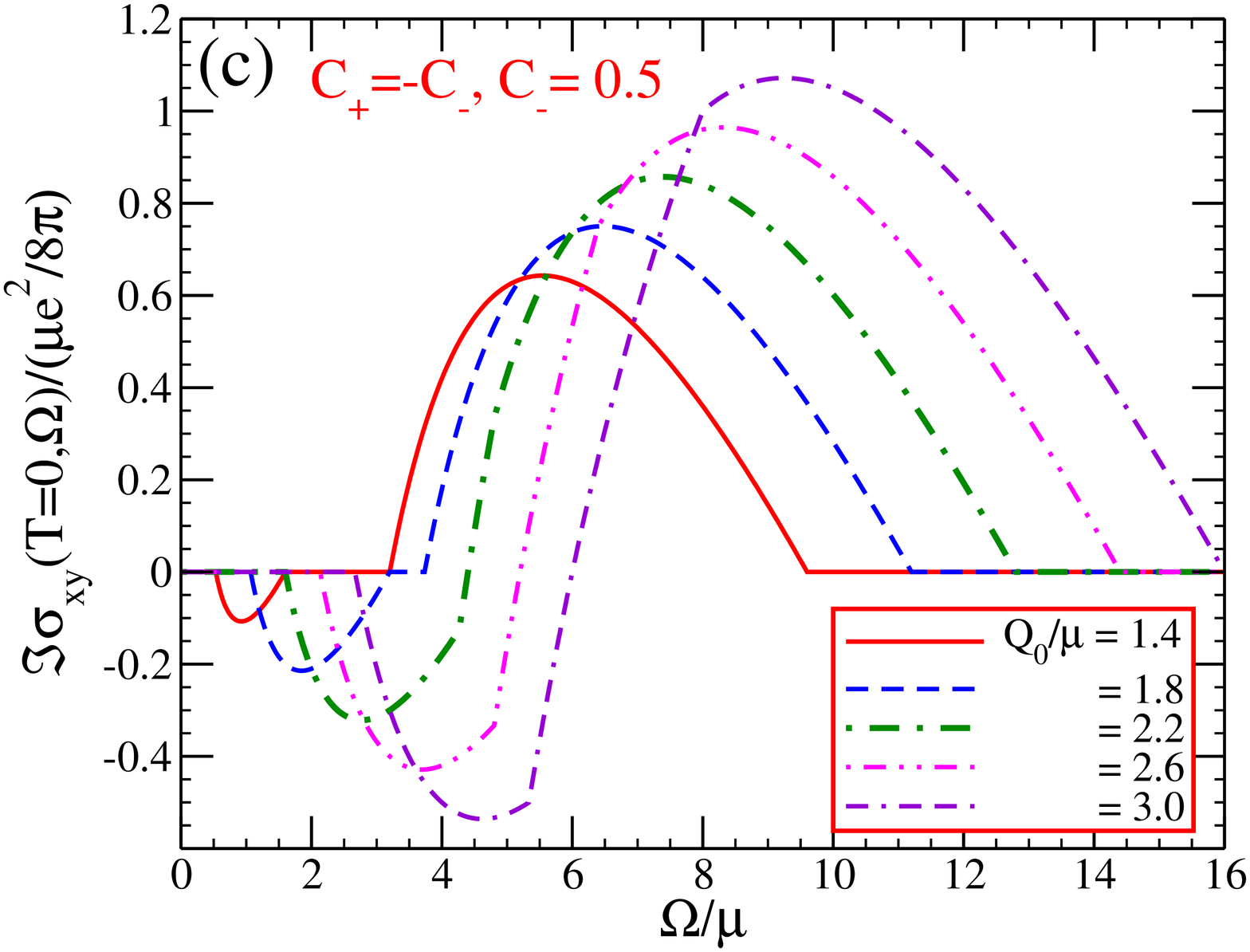}
\caption{(Color online)(a) Extended version of the phase diagram of Fig.(\ref{Fig2}a) for $\tQ_{0}>1$ with the contribution to the absorptive Hall conductivity indicated in the various distinct regions 
of the ($\tOmega,\tQ_{0}$) plane. `Null' indicates the regions of no absorptive Hall. The value of both the tilts are assumed to be $0.5$. In the other two sub-figures we plot the imaginary 
(absorptive) part of the AC Hall conductivity in units of $e^2 \mu/8\pi$ as a function of photon energy $\tOmega$, normalized to the chemical potential. There is a family of curves depending on the 
parameter $\tQ_{0}=Q_{0}/\mu$ which here is greater than one. Frame (b) is for $C_{+}=C_{-}$ while frame (c) has $C_{+}=-C_{-}$. This positive chirality Weyl node has reversed its tilt from counterclockwise 
to clockwise between (a) and (b) and this changes the sign of the positive chirality contribution to $\Im{\sigma_{xy}}(\Omega)$.} 
\label{Fig4}
\end{figure}

Next we consider the case of $\tQ_{0}>1$ staying with $C_{-}=0.5$. The relevant part of the phase diagram is shown in top frame (a) of Fig.(\ref{Fig4}). For $1<\tQ_{0}<2$ there is a region of zero 
absorption between $\tOmega^{-}_{U}$ and $\tOmega^{+}_{L}$. In the region $\tOmega^{-}_{L}$ to $\tOmega^{-}_{U}$ only the negative chirality node absorbs while in the region $\tOmega^{+}_{L}$ to 
$\tOmega^{+}_{U}$ it is only the positive chirality. However for $\tQ_{0}>2$ there is a overlap region between the $+$ and $-$ chirality contributions. There is no absorption below $\tOmega^{-}_{L}$ and 
above $\tOmega^{+}_{U}$. Between $\tOmega^{-}_{L}$ and $\tOmega^{+}_{L}$ only $\chi=-1$ is involved and in the region between $\tOmega^{-}_{U}$ and $\tOmega^{+}_{U}$ it is only $\chi=+1$ but in the 
intermediate region between 
\noindent
$\tOmega^{+}_{L}$ and $\tOmega^{-}_{U}$ there is overlap of the contributions to $\Im\sigma_{xy}(\Omega)$ 
of both chirality nodes. Detail plots of $\Im\sigma_{xy}(\Omega)$ vs. $\tOmega=\Omega/\mu$ are also provided in Fig.(\ref{Fig4}). Frame (b) is 
\noindent 
for $C_{+}=C_{-}$ and frame (c) $C_{+}=-C_{-}$ so that the 
contribution from the positive chirality node has reversed sign in frame (c) relative to (b). Further in frame (b) we note that $\Im{\sigma_{xy}}(\Omega)$ is always negative. The reason for this is that 
now the effective chemical potential for the negative chirality node has become negative and hence there has been a change of sign of this contribution relative to the case in Fig.(\ref{Fig3}) frame (b) 
where the contribution for the negative chirality node is positive. Finally in frame (c) of Fig.(\ref{Fig4}) we see the reversal in sign of the positive chirality
contribution because the sign of $C_{+}$ has been reversed in comparison to frame (b) of the same figure. For the solid red ($\tQ_{0}=1.4$) and dash blue ($\tQ_{0}=1.8$) curves the low energy region has 
$\sigma_{+}>\sigma_{-}$, the intermediate $\Omega$ region (no overlap) has $\sigma_{+}=\sigma_{-}$ and the region beyond this again has $\sigma_{+}>\sigma_{-}$ for parallel tilts and $\sigma_{+}<\sigma_{-}$ 
for opposite orientation of the tilts.

\section{Results for type II}
\label{sec:V}

\begin{figure}[H]
\centering
\includegraphics[width=1.0in,height=1.2in, angle=0]{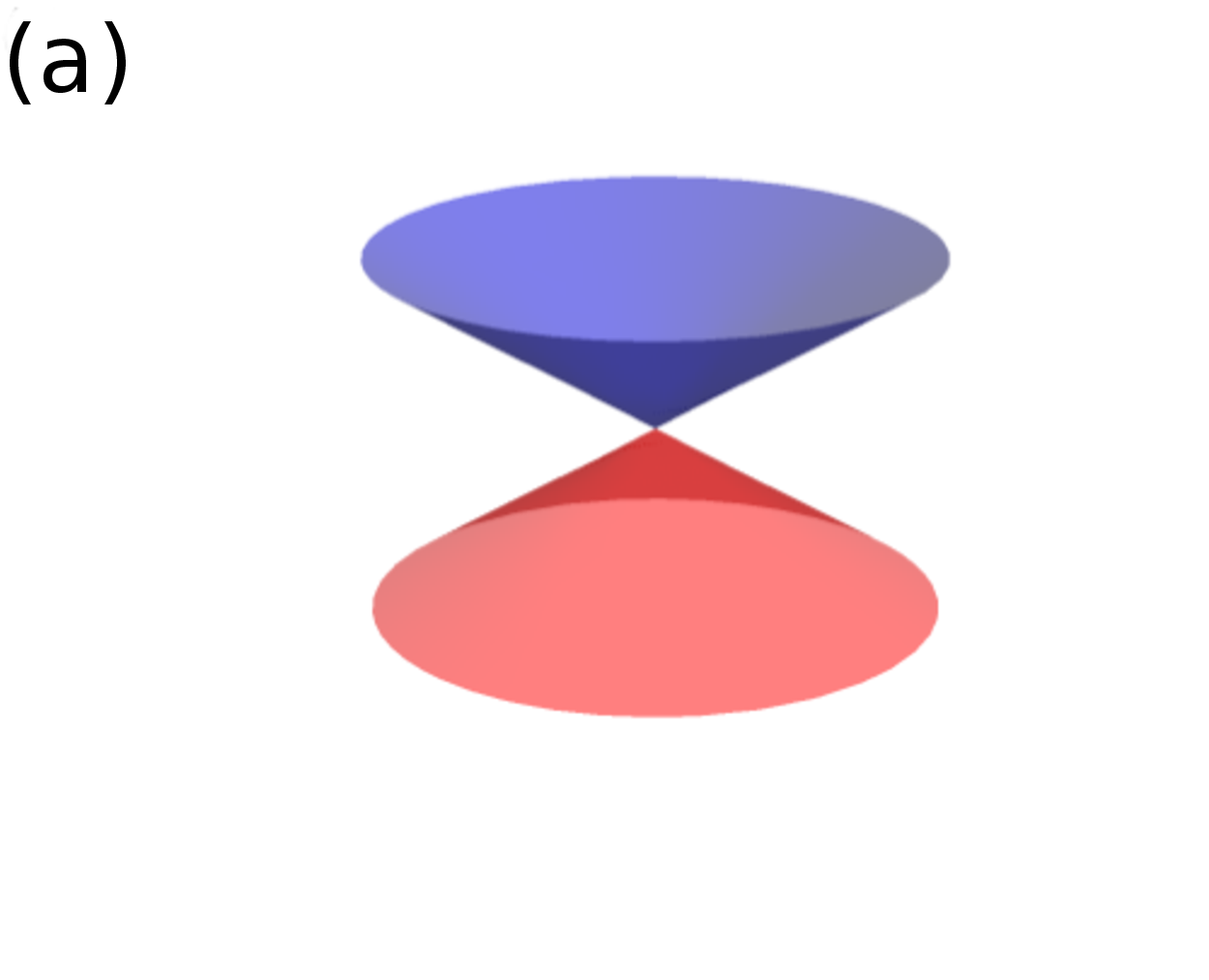}
\includegraphics[width=1.0in,height=1.2in, angle=0]{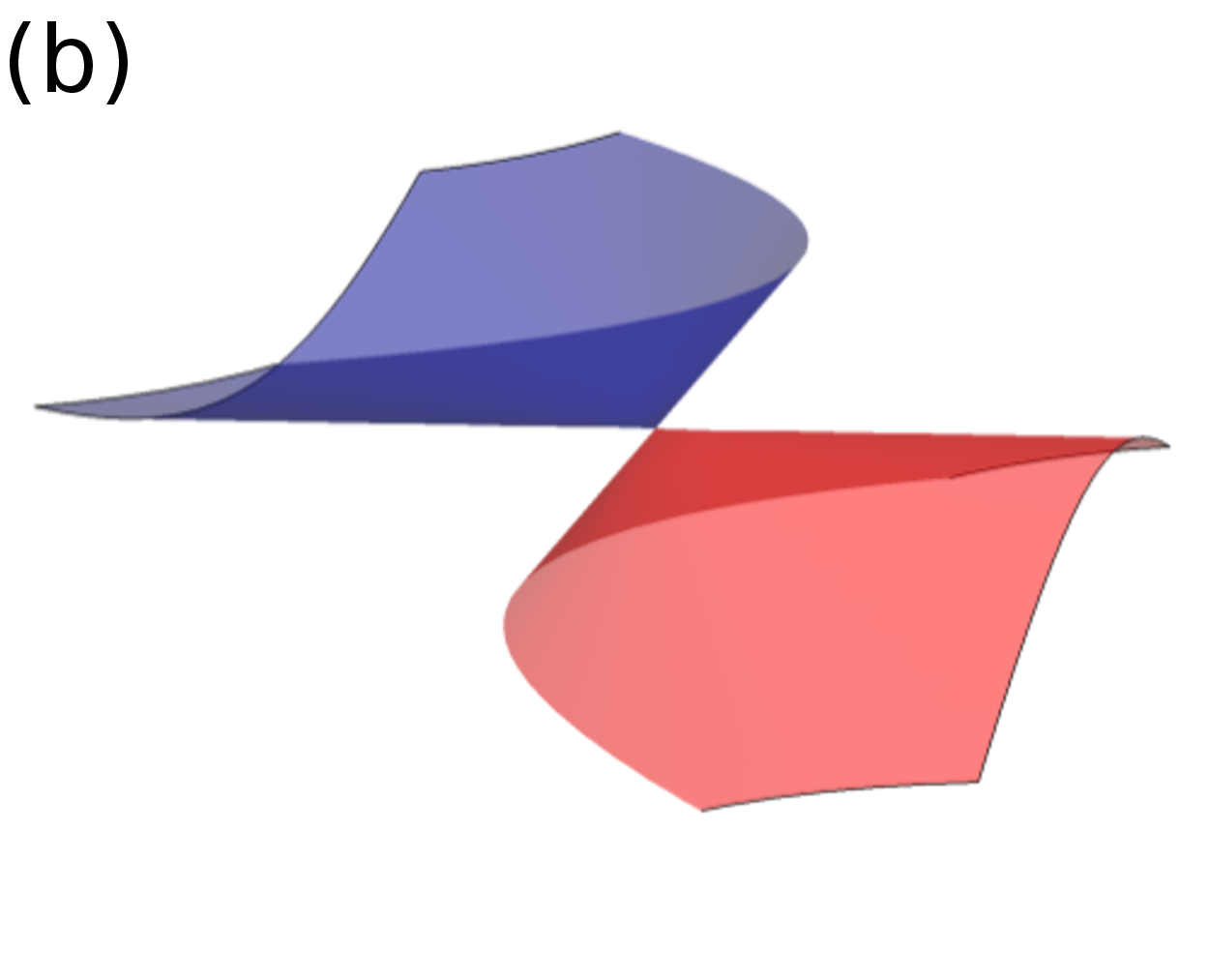}
\includegraphics[width=1.0in,height=1.2in, angle=0]{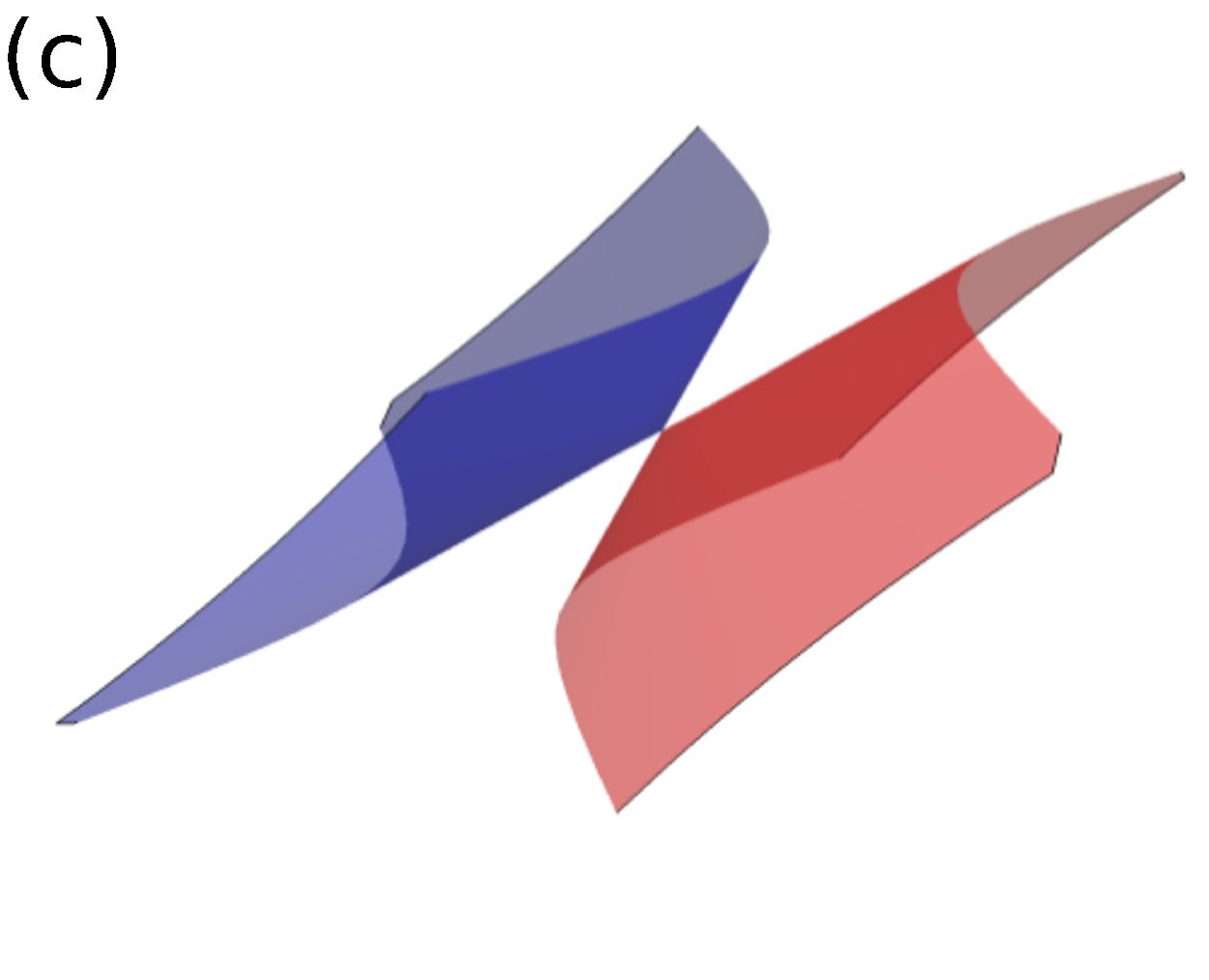}
\caption{(Color online) Schematic of tilted Weyl cones (a) $\bC=0$ no tilt, (b) $\bC=1$ tipped and (c) $\bC=2$ overtilted. Note that with increasing tilt the cross section of the Weyl cones increases as 
well in our model electronic dispersion curves based on the Hamiltonian of Eq.(\ref{Hamiltonian}).} 
\label{Fig5}
\end{figure}
\begin{figure}[H]
\centering
\includegraphics[width=2.0in,height=2.0in, angle=0]{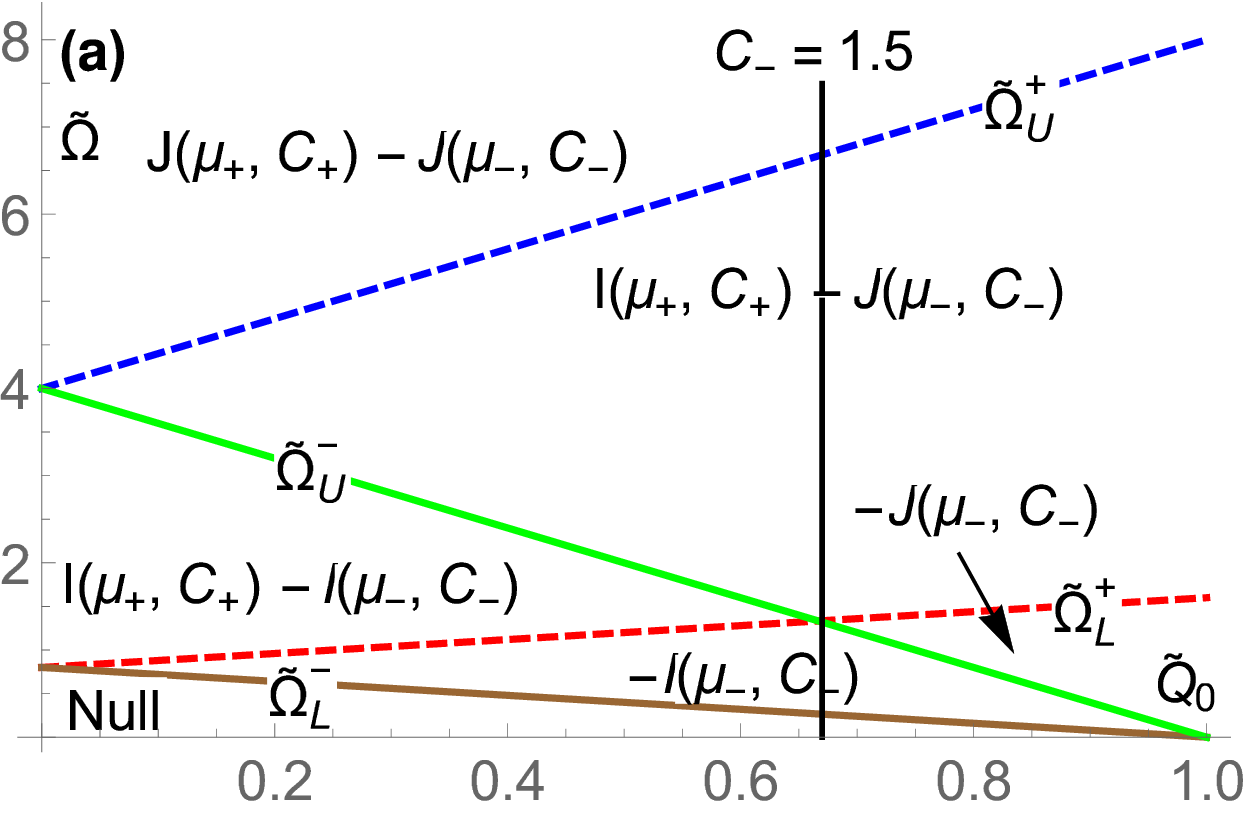}
\includegraphics[width=2.0in,height=2.5in, angle=270]{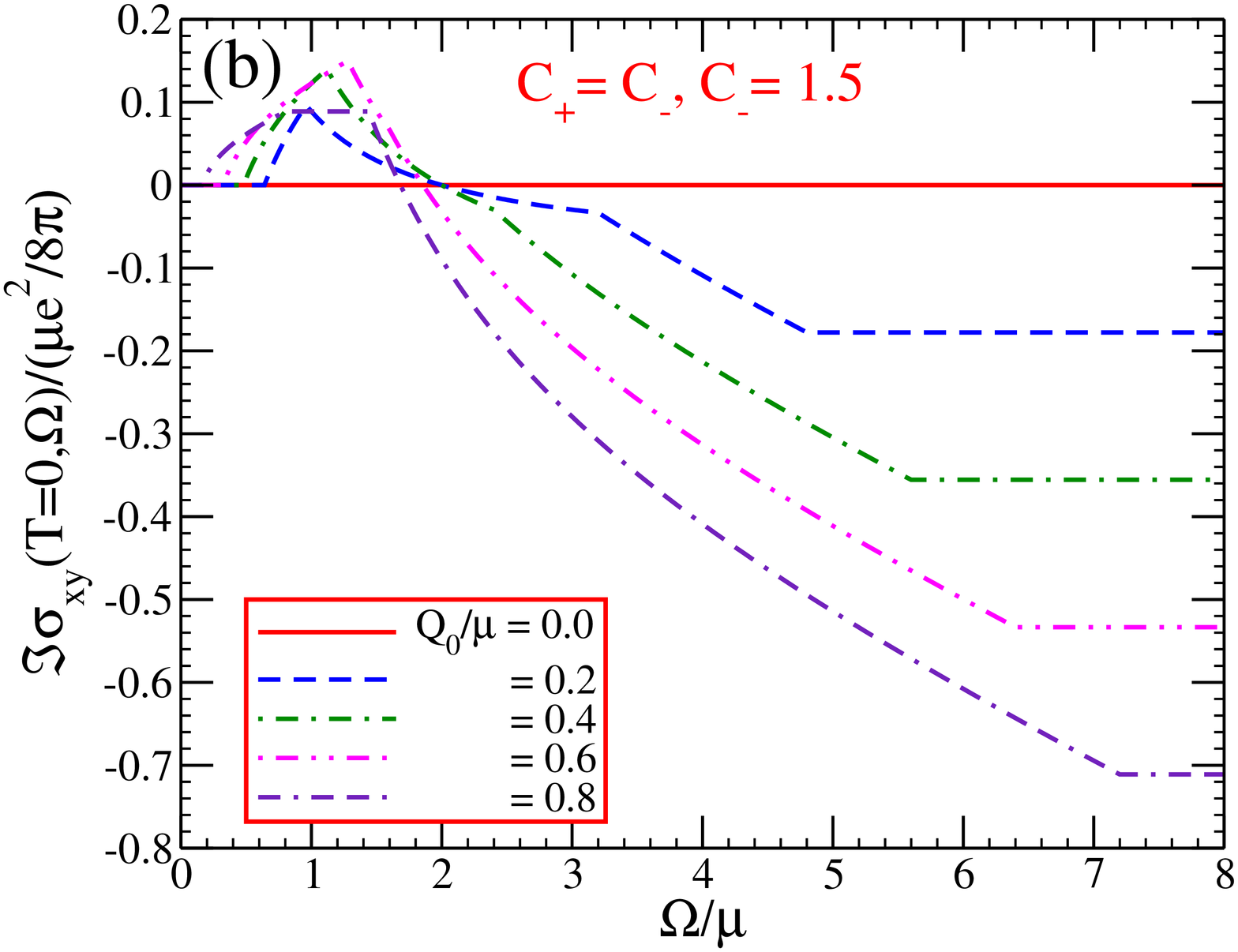}
\includegraphics[width=2.0in,height=2.5in, angle=270]{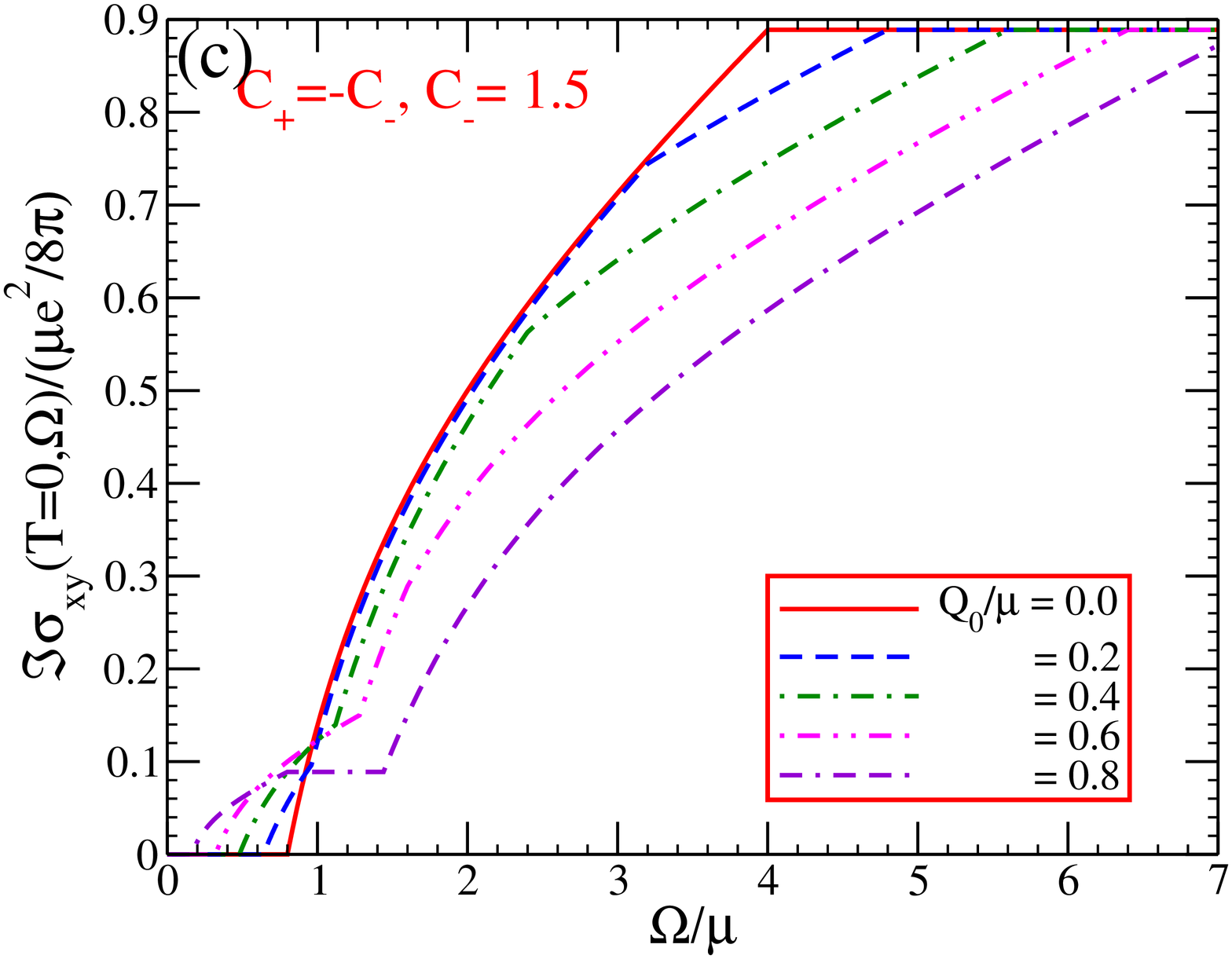}
\caption{(Color online) (a) Extended version of the phase diagram of Fig.(\ref{Fig2}a) for $\tQ_{0}<1$ with the contribution to the absorptive Hall conductivity indicated in the various distinct regions 
of the ($\tOmega, \tQ_{0}$) plane. `Null' indicates the regions of no absorptive Hall. The value of both the tilts are assumed to be $1.5$. In the other two sub-figures we plot the imaginary 
(absorptive) part of the AC Hall conductivity in units of $e^2 \mu/8\pi$ as a function of photon energy $\tOmega$, normalized to the chemical potential. There is a family of curves labeled by the parameter 
$\tQ_{0}=Q_{0}/\mu$ which here is less than one. Frame (b) is for $C_{+}=C_{-}$, both cones are tilted counterclockwise while in frame (c) the positive chirality cone has its tilt reversed $C_{+}=-C_{-}$ 
and this changes the sign of its contribution to $\Im{\sigma_{xy}}(\Omega)$.} 
\label{Fig6}
\end{figure}

We next turn to the overtilted case $C_{+}>1$ (type II Weyl). In Fig.(\ref{Fig5}), we show a schematic of the electronic dispersion curves for three values of the tilt $\bC=0$, left(a), no tilt; $\bC=1$ 
just tipped over (boundary between type I and type II), middle (b) and $\bC=2$ right(c), overtilted. As the tilt $\bC$ is increased the cone is bent over and opens up, and a Lifshitz transition occurs at 
$\bC=1$. For charge neutrality the density of states at the Fermi surface remains zero for $\bC<1$ but is finite for $\bC>1$ because of the formation of electron and hole pockets. Its value is dependent 
on the cut off \cite{Tiwari, Carbotte1} and on the value of $\bC$. For the imaginary part of the anomalous Hall conductivity this leads to the very different behavior seen in equation Eq.(\ref{FuncIcmu}) 
and (\ref{FuncIcmu1}). Eq.(\ref{FuncIcmu1}) does not enter the discussion in type I Weyl. Now both contributions $\chi=\pm1$ are unbounded above and there will be no intermediate regions of photon energies 
where there is no absorption. For definiteness we take a specific value of $C_{-}$ namely, $C_{-}=1.5$ and start with the case $\tQ_{0}<1$ for which all effective chemical potentials are positive. The 
relevant phase diagram is presented in the top frame (a) of Fig.(\ref{Fig6}) where 
\noindent
the different regions are indicated as well as the contributions of the two nodes to $\Im\sigma_{xy}(\Omega)$. Detail results for the Hall conductivity are given in Fig.(\ref{Fig6}) middle frame (b) and 
lower frame (c). Frame (b) applies to the case of
\begin{figure}[H]
\centering
\includegraphics[width=2.0in,height=2.0in, angle=0]{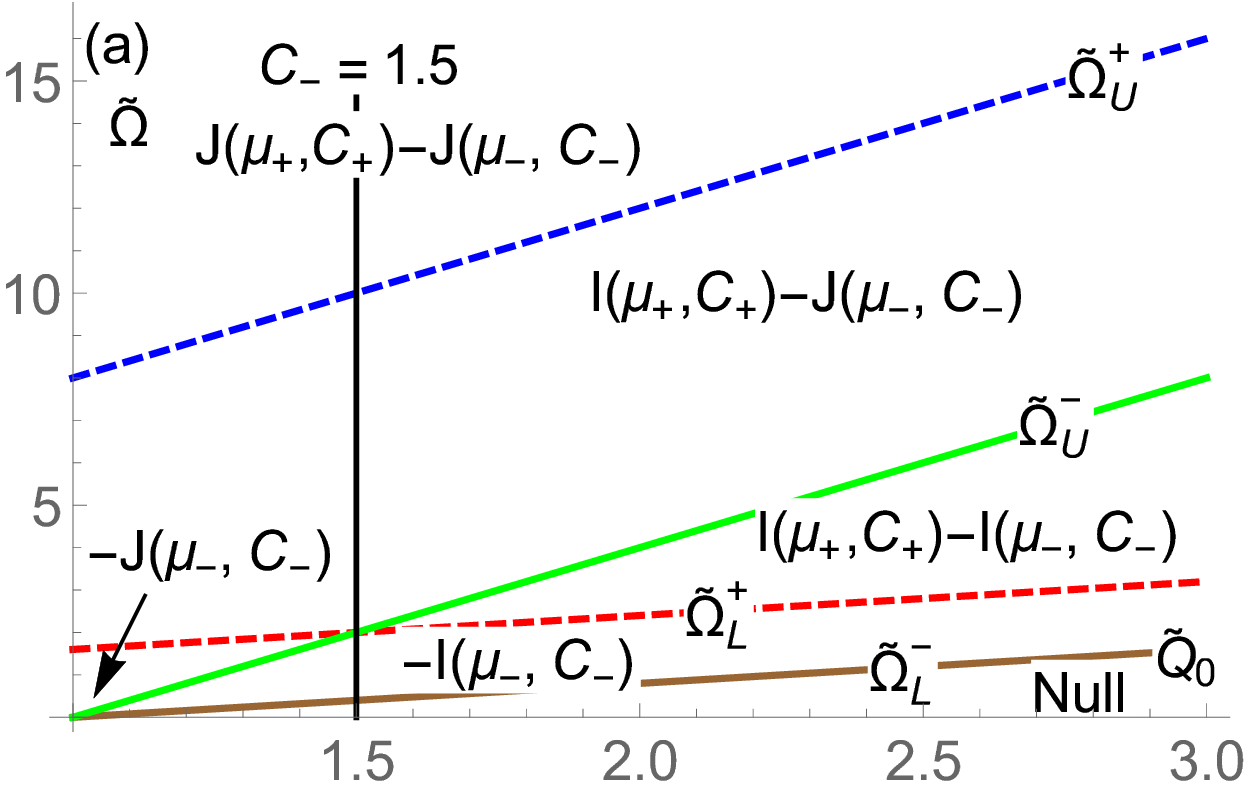}
\includegraphics[width=2.0in,height=2.5in, angle=270]{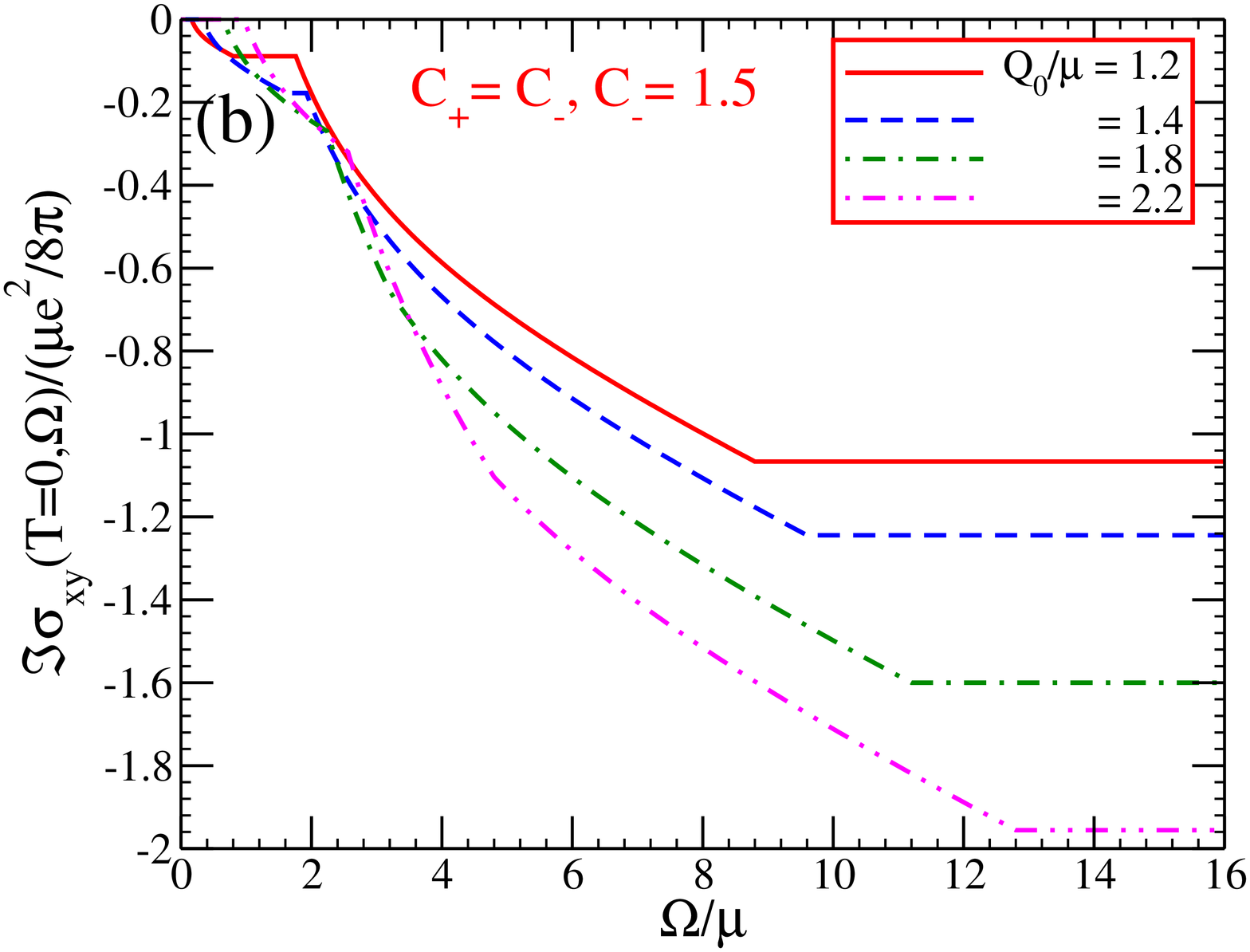}
\includegraphics[width=2.0in,height=2.5in, angle=270]{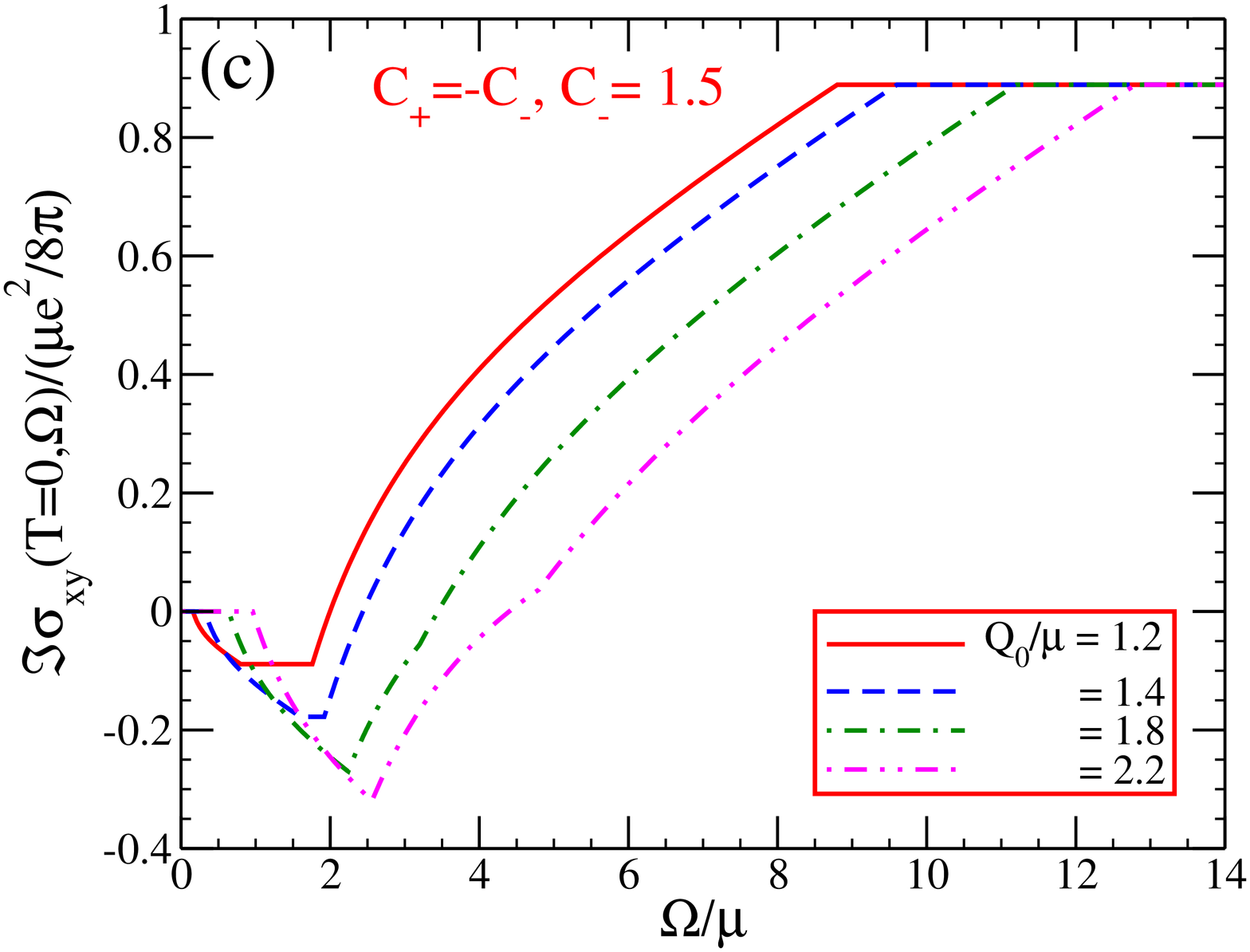}
\caption{(Color online)(a) Extended version of the phase diagram of Fig.(\ref{Fig2}a) for $\tQ_{0}>1$ with the contribution to the absorptive Hall conductivity indicated in the various distinct 
regions of the ($\tOmega, \tQ_{0}$) plane. `Null' indicates the regions of no absorptive Hall. The value of both the tilts are assumed to be $1.5$. In the other two sub-figures we plot the imaginary 
(absorptive) part of the AC Hall conductivity in units of $e^2 \mu/8\pi$ as a function of photon energy $\tOmega$, normalized to the chemical potential. There is a family of curves labeled by 
$\tQ_{0}=Q_{0}/\mu$ which here is greater than one. Frame (a) is for $C_{+}=C_{-}$, both cones are tilted counterclockwise while in frame (b) we have reversed the tilt of the positive chirality node to 
clockwise ($C_{+}=-C_{-}$) and this has changed the sign of its contribution to $\Im{\sigma_{xy}}(\Omega)$.} 
\label{Fig7}
\end{figure}
\noindent
an inversion symmetry breaking tilt while 
frame (c) is for tilts that respect inversion symmetry. As before we note that $\Im\sigma_{xy}(\Omega)$ is negative at high energies and positive at small energies for $C_{+}=C_{-}$ while for $C_{+}=-C_{-}$ 
it is everywhere positive because changing the tilt on the positive chirality node has changed the sign of its contribution. Two regions are worth special comments. As is clear from the phase diagram 
Fig.(\ref{Fig6})(a) for $\tQ_{0}>1/C_{-}\simeq 0.67$ there is an intermediate energy region between $\tOmega^{-}_{U}$ and $\tOmega^{+}_{L}$ where the absorption is equal to $-J(\mu_{-},C_{-})$ and hence 
is completely flat and due entirely to the negative chirality node. There is another flat region for any value of $\tQ_{0}<1$ when $\Omega>\tOmega^{+}_{U}$ and here both Weyl nodes contribute. The 
intermediate energy flat region is seen in the dot double dash indigo curve (Fig.(\ref{Fig6})frame (b)) which has $\tQ_{0}=0.8$ and this is greater than the critical value of $1/C_{-}$. In this region 
$\sigma_{+}$ and $\sigma_{-}$ will differ by a constant amount equal to $2|1-Q_{0}|/C^{2}_{-}$ in our conductivity units. In Fig.(\ref{Fig6}) frame (c) $C_{+}=-C_{-}$ and the sign of the contribution to 
$\Im{\sigma_{xy}}(\Omega)$ of the positive chirality node has switched from negative in frame (b) to positive because of the switch in the sign of the tilt. 

In Fig.(\ref{Fig7}) frame (a) we show the various regions of our phase diagram for $C_{-}=1.5$ and $\tQ_{0}>1$. In this case the effective chemical potential for the negative chirality node is negative. 
Note that for $\tQ_{0}<1.5$ there is a region between $\tOmega^{-}_{U}$ and $\tOmega^{+}_{L}$ where the Hall conductivity is determined by $-J(\mu_{-},C_{-})$ which corresponds to another region where 
$\Im{\sigma_{xy}}(\Omega)$ is flat but now the sign of this contribution is positive because there is one minus sign due to chirality and another due to a negative chemical potential. In the corresponding 
region of Fig.(\ref{Fig3}) $-I(\mu_{-},C_{-})$ is positive because $\mu_{-}$ is positive (see Fig.(\ref{Fig3}) frame (a)). This is seen in the solid red curve for $\tQ_{0}=1.2$ and is also part of the 
dash blue curve for $\tQ_{0}=1.4$ but in this case the photon energy range over which the flat contribution exist is very small because $\tOmega^{+}_{L}$ and $\tOmega^{-}_{U}$ are nearly the same in 
magnitude as they meet when $\tQ_{0}=1.5$. Finally frame (c) differs from frame (b) because of a change in sign of the positive chirality contribution.

\section{Discussion and conclusions}
\label{sec:VI}

We consider a Weyl semimetal with winding number one. Beside the usual relativistic Dirac Hamiltonian with degeneracy two, we further include in our work a term which explicitly breaks time reversal 
invariance and a second term which breaks inversion symmetry. Either of these two additional pieces lift the degeneracy of the Dirac cones and produces two Weyl nodes of opposite chirality. Violation of 
time reversal symmetry shifts the nodes in momentum space by $\pm \bf{Q}$ while inversion symmetry shifts them in energy by $\pm Q_{0}$. Using a Kubo formula for the absorptive part of the AC anomalous 
Hall conductivity ($\sigma_{xy}(\Omega)$) we compute the appropriate transverse current-current correlation function to get $\Im{\sigma_{xy}}(\Omega)$ vs. $\Omega$. We find that $\bf{Q}$ drops out of the 
expressions for the Hall conductivity in the clean limit employed here. Only the interband optical background is considered. In the limit of a centrosymmetric semimetal we recover the known expressions 
obtained by Steiner et. al [\onlinecite{Steiner}] for type I Weyl and by Mukherjee et. al [\onlinecite{Mukherjee1}] for both type I and II which they obtained within the context of a discussion of the 
absorption of circular polarized light. The two Weyl nodes contribute equally to $\Im{\sigma_{xy}}(\Omega)$ when they have opposite tilt and cancel when their tilts are parallel. For the opposite tilt 
($C_{-}$) and type I ($0<C_{-}<1$) the absorption is confined to a finite interval of photon energy $\tOmega\equiv\Omega/\mu$ between $2/(1+C_{-})$ to $2/(1-C_{-})$. For type II Weyl ($C_{-}>1$) the lower 
limit on $\tOmega$ for absorption remains but there is no upper limit and for $\bOmega>2/(C_{-}-1)$ the $\Im{\sigma_{xy}}(\Omega)$ in units of $\frac{e^2}{8\pi} \frac{\mu}{v}$ is a constant equal to 
$2/C^2_{-}$. 

For noncentrosymmetric Weyl the nodes are displaced in energy by $\pm Q_{0}$ and the effective chemical potential associated with each of the two nodes $\mu_{s'}=\mu+s'Q_{0}$ are different for $s'$, 
positive/negative chirality. The effective $\mu_{s'}$ for the negative chirality node is always smaller than for the positive node and becomes negative for $Q_{0}>\mu$ while $\mu_{+}$ is always positive 
for $\mu\ge0$ which has been assumed throughout this paper. The contribution to the absorptive part of the anomalous Hall conductivity of a given Weyl node was found to depend only on the magnitude of 
its effective chemical potential and tilt but its overall sign changes when the sign of $\mu_{s'}$ changes as it does when the sign of its tilt changes. The onset for absorption is different for each of 
the two nodes and equal to $\Omega=2|\mu_{s'}|/(1+C_{-})$. This means that for $Q_{0}\ne 0$ there will always be a low photon energy region for which only the negative chirality node contributes to the 
absorption. This is different from the $Q_{0}=0$ (centrosymmetric) case for which both nodes always contribute equally. For type I there are other regions of the phase diagram for $\tOmega$ vs $\tQ_{0}$ 
where only the positive chirality node contributes to $\Im{\sigma_{xy}}(\Omega)$. While above $\Omega=2\mu_{+}/(1-C_{-})$, $\Im{\sigma_{xy}}(\Omega)=0$, there is another window where it is also zero. For $\tQ_{0}$ 
between $C_{-}$ and 1 the photon energy interval is $2\lp\frac{1-\tQ_{0}}{1-C_{-}}\rp<\tOmega<2\lp\frac{1+\tQ_{0}}{1+C_{-}}\rp$ and for $\tQ_{0}$ between 1 and $1/C_{-}$ it is 
$2\lp\frac{\tQ_{0}-1}{1-C_{-}}\rp<\tOmega<2\lp\frac{1+\tQ_{0}}{1+C_{-}}\rp$. In these regions the absorption of light does not depend on its polarization. Such a region will never exist for centrosymmetric 
systems\cite{Mukherjee1}. It depends on the inversion symmetry breaking parameter $Q_{0}$ being large enough to produce two separated regions of photon energies in which only one chirality Weyl node 
contributes. By contrast for type II Weyl there is always a finite value of $\Im{\sigma_{xy}}(\Omega)$ for  $\Omega>2\mu_{-}/(1+C_{-})$. For $\Omega>2\mu_{+}/(1-C_{-})$ the Hall conductivity in units of 
$\frac{e^2}{8\pi} \frac{\mu}{v}$ takes on a particularly simple form. For parallel tilt it is a constant equal to $-\tQ_{0}/C^2_{-}$ while for the opposite tilt it is again constant but equal to 
$-2/C^2_{-}$. In other regions of the phase diagram for both type I and type II and arbitrary direction of tilt, photon energy the Hall conductivity $\Im{\sigma_{xy}}(\Omega)$ has a more complicated 
dependence but can still be expressed in terms of two simple algebraic functions $I(\bC,\bmu)$ and $J(\bC,\bmu)$ which we defined in Eq.(\ref{FuncIcmu}) and (\ref{FuncIcmu1}) respectively. These are one 
of our central results. For type II there is a second region closely related to that seen in type I where,  in that case, $\Im{\sigma_{xy}}(\Omega)$ was zero. Now it is instead constant and entirely due 
to the negative Weyl node. For $\tQ_{0}$ between $1/C_{-}$ to 1 it is $\frac{1-\tQ_{0}}{C^2_{-}}$ in units of $\frac{e^2}{8\pi}\frac{\mu}{v}$ in the photon energy interval
$2\lp\frac{1-\tQ_{0}}{C_{-}-1}\rp<\tOmega<2\lp\frac{1+\tQ_{0}}{C_{-}+1}\rp$. For $\tQ_{0}$ between 1 to $C_{-}$ the interval of constant $\Im{\sigma_{xy}}(\Omega)$ is 
$2\lp\frac{\tQ_{0}-1}{C_{-}-1}\rp<\tOmega<2\lp\frac{1+\tQ_{0}}{C_{-}+1}\rp$. In these two regions the absorptive part of the conductivity associated with circular polarized light $\sigma_{+}$ and 
$\sigma_{-}$ differ only by a constant amount $\frac{2|1-\tQ_{0}|}{C^{2}_{-}}$ in our units. 

In summary we have found a rich and complex phase diagram for the variation of the imaginary part of the AC Hall conductivity as a function of the inversion symmetry breaking parameter $Q_{0}$, doping 
$\mu$ and Weyl cone tilt $C$ in noncentrosymmetric Weyl semimetals. Our result also impact on the absorption of circular polarized light. For a pair of opposite chirality Weyl nodes tilts that preserve 
inversion symmetry (oppositely tilted), have a different signature in optics than do tilts that do not (parallel tilts).

\subsection*{Acknowledgments}
Work supported in part by the Natural Sciences and Engineering Research Council of Canada (NSERC)(Canada) and by the Canadian Institute for Advanced Research (CIFAR)(Canada). We thanks A. A. Burkov and 
D. Xiao for enlightening correspondence.

\end{document}